\documentclass{pasj00}
\draft

\begin{document}
\SetRunningHead{S. Hasegawa et al.}{The Opposition Surge of Vesta}

\title{The Opposition Effect of the Asteroid 4 Vesta}

\author{%
   Sunao \textsc{Hasegawa},\altaffilmark{1}
   Seidai \textsc{Miyasaka},\altaffilmark{2}
   Noritaka \textsc{Tokimasa},\altaffilmark{3}
   Akito \textsc{Sogame},\altaffilmark{4}
   Mansur A. \textsc{Ibrahimov},\altaffilmark{5}
   Fumi \textsc{Yoshida},\altaffilmark{6}
   Shinobu \textsc{Ozaki},\altaffilmark{7}
   Masanao \textsc{Abe},\altaffilmark{1}
   Masateru \textsc{Ishiguro},\altaffilmark{8}
   and 
   Daisuke \textsc{Kuroda}\altaffilmark{9}
}
 \altaffiltext{1}{Institute of Space and Astronautical Science, Japan Aerospace Exploration Agency, 3-1-1 Yoshinodai, Chuo-ku, Sagamihara, Kanagawa 252-5210, Japan}
 \email{hasehase@isas.jaxa.jp}
 \altaffiltext{2}{Tokyo Metropolitan Government, 2-8-1 Nishishinjyuku, Shinjyuku, Tokyo 163-8001, Japan}
 \altaffiltext{3}{Sayo Town Office, 2611-1 Sayo, Sayo-cho, Sayo, Hyogo 679-5380, Japan}
 \altaffiltext{4}{School of Engineering, Tokai University, 4-1-1 Kitakaname, Hiratsuka, Kanagawa 259-1292, Japan}
 \altaffiltext{5}{Space Astrometry Department, Institute of Astronomy of the Russian Academy of Sciences, 48 Pyatnitskaya Street 119017, Moscow, Russia}
 \altaffiltext{6}{Office of International Relations, National Astronomical Observatory of Japan, 2-21-1 Osawa, Mitaka, Tokyo 181-8588, Japan}
 \altaffiltext{7}{TMT Project Office, National Astronomical Observatory of Japan, 2-21-1 Osawa, Mitaka, Tokyo 181-8588, Japan}
 \altaffiltext{8}{Department of Physical and Astronomy, Seoul National University, 1 Gwanak-ro, Gwanak-gu, Seoul 151-742, Korea}
 \altaffiltext{9}{Okayama Astrophysical Observatory, National Astronomical Observatory of Japan, 3037-5 Honjo, Kamogata-cho, Asakuchi, Okayama 719-0232, Japan}


%

\KeyWords{Planetary Systems  ---  minor planets, asteroids: individual (Vesta)  ---  minor planets, asteroids: general ---  techniques:photometric} 

\maketitle

\begin{abstract}
We present the results of photometric observations carried out with four small telescopes of the asteroid 4 Vesta in the $B$, $R_{\rm C}$, and $z'$ bands at a minimum phase angle of 0.1$\timeform{D}$.
The magnitudes, reduced to unit distance and zero phase angle, were $M_{B}(1, 1, 0) = 3.83 \pm 0.01, M_{R_{\rm C}}(1, 1, 0)  = 2.67 \pm 0.01$, and $M_{z'}(1, 1, 0) = 3.03 \pm 0.01$ mag.
The absolute magnitude obtained from the IAU $H$--$G$ function is $\sim$0.1 mag darker than the magnitude at a phase angle of 0$\timeform{D}$ determined from the Shevchenko function and Hapke models with the coherent backscattering effect term.
Our photometric measurements allowed us to derive geometric albedos of 0.35 in the $B$ band, 0.41 in the $R_{\rm C}$ band, and 0.31 in the $z'$ bands by using the Hapke model with the coherent backscattering effect term.
Using the Hapke model, the porosity of the optically active regolith on Vesta was estimated to be $\rho$ = 0.4--0.7, yielding the bluk density of 0.9--2.0 $\times$ $10^3$ kg $\mathrm{m^{-3}}$.
It is evident that the opposition effect for Vesta makes a contribution to not only the shadow-hiding effect, but also the coherent backscattering effect that appears from ca. $1\timeform{D}$.
The amplitude of the coherent backscatter opposition effect for Vesta increases with a brightening of reflectance. 
By comparison with other solar system bodies, we suggest that multiple-scattering on an optically active scale may contribute to the amplitude of the coherent backscatter opposition effect ($B_{C0}$).

\end{abstract}

\section{Introduction}
\label{sec:Introduction}
The asteroid 4 Vesta was the last of four asteroids first discovered by H.W. Olbers in 1807.
Vesta is the brightest of the main-belt asteroids observed from Earth, and has been the subject of numerous observational studies.

The diameter of Vesta was determined in the 1820s \citep{Aspin1825} using a micrometrical measure.
The albedo value and brightness variation of Vesta were also both measured prior to 1990 \citep{Harrington1883}.
\citet{Bobrovnikoff1929} first determined the rotational periods and spectrum of Vesta.
Evaluation of the perturbations due to Vesta during an asteroid approach near Vesta allowed derivation of the mass of Vesta \citep{Hertz1968}.
A radiometric technique was first applied to determine an accurate albedo and diameter for an asteroid (Vesta) by \citet{Allen1970}.
\citet{McCord1970} showed that certain basaltic achondrite meteorites can be linked by visible reflectance measurements to Vesta.
Using lightcurves for Vesta, two possible solutions of pole orientation and the existence of a south pole crater were proposed \citep{Taylor1973}.
\citet{Larson1975} concluded, based on spectroscopic data obtained at near-infrared wavelengths, that the Vestan surface is covered with eucrites, which are part of the howardite--eucrite--diogenite (HED) achondrite meteorite clan.
Disk resolved images of Vesta have been derived from speckle interferometry techniques \citep{Worden1977}.
\citet{Degewij1979} monitored Vesta polarimetrically and concluded that it has a nearly spheroidal shape with a heterogeneous surface.
Ultraviolet spectra were obtained for Vesta using the International Ultraviolet Explorer (IUE: \cite{Boggess1978}), and the results supported the link between Vesta and HED meteorites \citep{Butterworth1980}.
\citet{Drummond1988} delineated surface patterns on Vesta from speckle interferometric observations.
Centimeter and millimeter observations indicate the presence of dusty regolith on the surface of Vesta \citep{Johnston1989}.
From radar albedo and the circular polarization ratio obtained using radar observations, Vesta has been shown to have a basaltic and immature regolith surface \citep{Mitchell1996}.
Disk resolved images with the Hubble Space Telescope (HST) allowed surface color variations and topographic features of Vesta to be mapped in detail \citep{Thomas1997}.
\citet{Mueller1998} conducted a thermal inertia study of the surface of Vesta using a combination of thermal radiation data from the Infrared Astronomical Satellite (IRAS: \cite{Neugebauer1984}), Infrared Space Observatory (ISO: \cite{Kessler1996}), and ground-based observatories.
\citet{Dotto2000} and \citet{Heras2000} noted that olivine-associated spectral features are evident on the surface of Vesta using an ISO Photo-Polarimeter and Short Wavelength Spectrometer, respectively. 
The detection of a hydrated absorption feature in Earth-based KL band measurements has been interpreted to represent carbonaceous chondritic ejecta and space weathering on Vesta \citep{Hasegawa2003}.
Prior to the Dawn spacecraft's rendezvous with Vesta \citep{Russell2012}, remote sensing measurements using ground- and space-based observatories had revealed many aspects of the physical properties of Vesta, as described above.

The opposition effect is the brightening of a particulate medium back in the direction toward a light source, and is also known as the non-linear surge at small phase angles. 
The opposition effect was first observed for Saturnian rings by \citet{Seeliger1895}, for the Moon by \citet{Barabashev1922}, for Mars by \citet{O'Leary1967}, and for Iapetus by \citet{Franklin1974}. 
The opposition effect from an asteroid was recognized for the first time by \citet{Gehrels1956}.
\citet{Gehrels1977} compared the opposition effects of five asteroids.
Compiling previous data at phase angles of less than 0.3$\timeform{D}$, \citet{Belskaya2000} studied the opposition effect for 33 asteroids comprising various spectral types.
Characteristics of the opposition effect of 21 dark asteroids were obtained down to phase angles of 0.1$\timeform{D}$--0.9$\timeform{D}$ (\authorcite{Shevchenko1997} \yearcite{Shevchenko1997}, \yearcite{Shevchenko2002}, \yearcite{Shevchenko2008}, \yearcite{Shevchenko2012}).

The shadow-hiding and coherent backscattering enhancement mechanisms are considered to be major contributors to the opposition effect.
For high-albedo objects, such as E-type asteroids \citep{Harris1989} with a mean geometric albedo of 0.55 $\pm$ 0.21 (\authorcite{Usui2013} \yearcite{Usui2013}, \yearcite{Usui2014}) and Galilean satellites \citep{Thompson1992} with a high albedo of $\sim$0.8 \citep{Buratti1995}, the point where the brightness changes from linear growth to a steeply sloped surge is in the range of several degrees. 
The opposition effect for higher-albedo surfaces is generally explained by coherent backscattering enhancement (\cite{Muinonen2002}, \cite{Dlugach2013}).
\citet{Kaydash2013} proposed that coherent backscattering enhancement partially contributes to the opposition effect of the Moon, which has a medium albedo that is lower than that of Vesta at phase angles less than 2$\timeform{D}$.
Some dark asteroids display a large opposition effect over a range of phase angles less than 1$\timeform{D}$ \citep{Belskaya2002}.
Therefore, observations at low phase angles (i.e., less than 1$\timeform{D}$) are important to constrain the properties of the opposition effect for asteroids.

The opposition effect for Vesta was first detected by \citet{Gehrels1967}.
Photometric observations of Vesta, including at small phase angles less than several degrees, have been carried out in numerous studies (1.9$\timeform{D}$; \cite{Gehrels1967}, 2.5$\timeform{D}$; \cite{Taylor1973}, 1.4$\timeform{D}$; \cite{Lagerkvist1989}, 1.7$\timeform{D}$; \cite{Lagerkvist1990}, 1.3$\timeform{D}$; \cite{Rock1990}, 1.2$\timeform{D}$; \cite{Lagerkvist1992}).
The Dawn spacecraft obtained photometric images whilst orbiting 4 Vesta at phase angles from 7.7$\timeform{D}$ to 107.5$\timeform{D}$ \citep{Li2013}.
However, photometric data for Vesta at less than 1$\timeform{D}$, which would allow a robust investigation of the opposition effect, have not yet been obtained.

On January 5 2006, the phase angle of Vesta was located at 0.1$\timeform{D}$, and such a small phase angle (less than 1.0$\timeform{D}$) will not happen again until 2029. 
To investigate the characteristics of the opposition effect and phase function at an extremely low phase angle for 4 Vesta, we obtained photometric data for Vesta from December 2005 to April 2006.
Here we present these photometric observations. 
This paper describes the observations and data reduction procedures (Section \ref{sec:Observations and data reduction procedures}), acquisition of the photometric data (Section \ref{sec:Photometric data}), phase function results (Section \ref{sec:Phase function}), and a discussion of these results (Section \ref{sec:Discussion}).

\section{Observations and data reduction procedures}
\label{sec:Observations and data reduction procedures}
The observations of Vesta were performed using six different telescopes at five locations in Japan and Uzbekistan, including photometry between December 2005 and April 2006, and spectroscopy during February and March 2006, and January 2014.
Photometric data were obtained at the Institute of Space and Astronautical Science (ISAS), Japan Aerospace Exploration Agency (JAXA) in Kanagawa, Japan (no MPC code; 139$\timeform{D}$23'43"E, 35$\timeform{D}$33'29"N; 140 m), Miyasaka Observatory in Yamanashi, Japan (MPC code 366; 138$\timeform{D}$17'50"E, 35$\timeform{D}$51'57"N; 860 m), Nishiharima Astronomical Observatory in Hyogo, Japan (no MPC code; 134$\timeform{D}$20'09"E, 35$\timeform{D}$01'33"N; 450 m), and Maidanak Astronomical Observatory in Qashqadaryo, Uzbekistan (no MPC code; 66$\timeform{D}$53'51"E, 38$\timeform{D}$40'25"N; 2600 m).
Spectroscopic data were recorded at the Nishiharima Astronomical Observatory and the Okayama Astrophysical Observatory, National Astronomical Observatory of Japan in Okayama, Japan (MPC code 371; 133$\timeform{D}$35'38"E, 34$\timeform{D}$34'37"N; 360 m).

\subsection{Photometric observations}
\label{sec:Photometric observations}
Photometric observations of Vesta were recorded for 30 nights.
The nightly observing details are summarized in Table \ref{tab:photometric circumstances}.

\begin{longtable}{ccccccc}
  \caption{Nightly details of the photometric observations.\footnotemark[$*$]}\label{tab:photometric circumstances}
  \hline
   Date & Time (UT) & $R_{\rm h}$  & $\Delta$  & $\alpha$ & Telescope & Filter
\\ 
   $\rm[year.month.day]$ & [hour.minute.second] &  [AU] &  [AU] &  [$\timeform{D}$] & & band
\\
\endfirsthead
  \hline
   Date & Time (UT) & $R_{\rm h}$ & $\Delta$ & $\alpha$ & Telescope & Band
\\
  \hline
\endhead
  \hline
\endfoot
  \hline
\multicolumn{1}{@{}l}{\rlap{\parbox[t]{1.0\textwidth}{\small
\footnotemark[$*$] The heliocentric distance ($R_{\rm h}$), geocentric distance ($\Delta$), and phase angle ($\alpha$) for observing asteroids were obtained by the JPL HORIZON ephemeris generator system of NASA.\footnotemark[1]
}}}
\endlastfoot
  \hline
2005.Dec.01&12:32:31 -- 18:34:31&2.552&1.741& 15.3&0.064 m -- Sagamihara&$BR_{\rm C}z'$\\
2005.Dec.08&12:37:18 -- 19:09:15&2.549&1.680&12.8 -- 12.9&0.064 m -- Sagamihara&$BR_{\rm C}z'$\\
2005.Dec.09&12:48:40 -- 20:02:04&2.549&1.672&12.4 -- 12.5&0.064 m -- Sagamihara&$BR_{\rm C}z'$\\
2005.Dec.22&11:47:41 -- 19:20:44&2.543&1.590&~6.8 -- ~7.0&0.064 m -- Sagamihara&$BR_{\rm C}z'$\\
2005.Dec.24&11:17:38 -- 18:44:57&2.542&1.581&~6.0 -- ~6.1&0.064 m -- Sagamihara&$BR_{\rm C}z'$\\
2005.Dec.28&11:30:15 -- 16:04:49&2.540&1.567&~4.1 -- ~4.2&0.064 m -- Sagamihara&$BR_{\rm C}z'$\\
2005.Dec.30&11:04:36 -- 17:53:25&2.539&1.562&~3.1 -- ~3.2&0.064 m -- Sagamihara&$BR_{\rm C}z'$\\
           &12:25:34 -- 19:01:15&2.539&1.562&~3.1 -- ~3.2&0.36 m -- Miyasaka&$BR_{\rm C}z'$\\
2005.Dec.31&13:03:20 -- 18:40:14&2.539&1.559&~2.6 -- ~2.7&0.36 m -- Miyasaka&$BR_{\rm C}z'$\\
2006.Jan.02&12:46:02 -- 19:49:22&2.538&1.556&~1.6 -- ~1.7&0.36 m -- Miyasaka&$BR_{\rm C}z'$\\
2006.Jan.03&10:52:58 -- 19:10:20&2.537&1.555&~1.1 -- ~1.3&0.064 m -- Sagamihara&$BR_{\rm C}z'$\\
           &11:34:44 -- 18:41:44&2.537&1.555&~1.1 -- ~1.2&0.36 m -- Miyasaka&$BR_{\rm C}z'$\\
           &12:49:43 -- 13:30:33&2.537&1.555& 1.2&0.076 m -- Nishiharima&$BR_{\rm C}z'$\\
2006.Jan.05&10:09:27 -- 17:12:14&2.536&1.553&~0.2 -- ~0.3&0.064 m -- Sagamihara&$BR_{\rm C}z'$\\
           &10:18:27 -- 20:15:35&2.536&1.553&~0.1 -- ~0.3&0.36 m -- Miyasaka&$BR_{\rm C}z'$\\
           &14:25:09 -- 16:05:04&2.536&1.553& 0.2&0.076 m -- Nishiharima&$BR_{\rm C}z'$\\
           &22:31:43 -- 23:28:47&2.536&1.553& 0.1&0.60 m -- Maidanak&$BR_{\rm C}$\\
2006.Jan.06&15:02:09 -- 19:32:53&2.535&1.552& 0.4&0.36 m -- Miyasaka&$BR_{\rm C}z'$\\
           &16:33:20 -- 16:59:32&2.535&1.552& 0.4&0.076 m -- Nishiharima&$BR_{\rm C}z'$\\
2006.Jan.07&10:26:21 -- 17:50:14&2.535&1.552&~0.8 -- ~0.9&0.064 m -- Sagamihara&$BR_{\rm C}z'$\\
           &10:42:44 -- 19:10:30&2.535&1.552&~0.8 -- ~0.9&0.36 m -- Miyasaka&$BR_{\rm C}z'$\\
2006.Jan.08&10:36:44 -- 14:33:54&2.535&1.552&~1.2 -- ~1.3&0.064 m -- Sagamihara&$BR_{\rm C}z'$\\
           &10:31:09 -- 16:55:31&2.535&1.552&~1.2 -- ~1.4&0.36 m -- Miyasaka&$BR_{\rm C}z'$\\
2006.Jan.25&10:20:59 -- 17:59:58&2.525&1.598&~9.4 -- ~9.5&0.064 m -- Sagamihara&$BR_{\rm C}z'$\\
2006.Jan.27&10:43:40 -- 17:37:28&2.524&1.608&10.2 -- 10.4&0.064 m -- Sagamihara&$BR_{\rm C}z'$\\
           &12:59:10 -- 13:48:06&2.524&1.608&10.3&0.36 m -- Miyasaka&$BR_{\rm C}z'$\\
2006.Jan.28&09:01:31 -- 15:18:34&2.523&1.613&10.6 -- 10.8&0.36 m -- Miyasaka&$BR_{\rm C}z'$\\
2006.Feb.04&09:22:49 -- 15:21:43&2.519&1.659&13.5 -- 13.6&0.36 m -- Miyasaka&$BR_{\rm C}z'$\\
2006.Feb.11&09:46:00 -- 15:43:07&2.515&1.715&16.0&0.36 m -- Miyasaka&$BR_{\rm C}z'$\\
2006.Feb.17&10:57:56 -- 16:31:04&2.511&1.770&17.8 -- 17.9&0.36 m -- Miyasaka&$BR_{\rm C}z'$\\
2006.Feb.18&12:50:01 -- 16:05:04&2.510&1.779&18.1&0.076 m -- Nishiharima&$BR_{\rm C}z'$\\
2006.Feb.23&10:57:25 -- 14:49:22&2.506&1.829&19.4&0.076 m -- Nishiharima&$BR_{\rm C}z'$\\
2006.Mar.03&09:29:08 -- 11:33:53&2.501&1.915&21.0 -- 21.1&0.36 m -- Miyasaka&$BR_{\rm C}z'$\\
2006.Mar.04&09:36:50 -- 11:44:23&2.501&1.927&21.2&0.36 m -- Miyasaka&$BR_{\rm C}z'$\\
2006.Mar.14&09:34:59 -- 15:09:59&2.493&2.044&22.6 -- 22.7&0.064 m -- Sagamihara&$BR_{\rm C}z'$\\
2006.Mar.15&09:36:09 -- 12:17:59&2.493&2.055&22.8&0.064 m -- Sagamihara&$BR_{\rm C}z'$\\
2006.Mar.20&09:42:59 -- 12:12:18&2.489&2.115&23.2&0.064 m -- Sagamihara&$BR_{\rm C}z'$\\
           &11:05:51 -- 13:42:03&2.489&2.116&23.2&0.076 m -- Nishiharima&$BR_{\rm C}z'$\\
2006.Mar.29&10:54:30 -- 13:53:58&2.482&2.225& 23.7&0.064 m -- Sagamihara&$BR_{\rm C}z'$\\
2006.Mar.31&09:43:42 -- 13:49:44&2.480&2.250&23.7 -- 23.8&0.064 m -- Sagamihara&$BR_{\rm C}z'$\\
2006.Apr.03&10:47:08 -- 13:33:19&2.478&2.287&23.8&0.076 m -- Nishiharima&$BR_{\rm C}z'$\\
\end{longtable}
\footnotetext[1]{$\langle$http://ssd.jpl.nasa.gov/horizons.cgi\#top$\rangle$.}

Eighteen nights of photometric observations were made with a refracting telescope with an aperture of 0.064 m and a focal ratio of 2.8 at ISAS, JAXA, in Sagamihara.
This telescope was installed temporarily for this study on the roof of the research/administration building.
An SBIG ST-10XME with a Kodak KAF-3200ME detector yielded a format of 2184 $\times$ 1472 pixels with an image scale of \timeform{7".8}/pixel and a sky field of \timeform{282'.6} $\times$ \timeform{190'.8}. 
The full-width half maximum of stellar images at this site was typically 3--5\timeform{"}.
The resolution is smaller than the pixel size, but star images were spread out over several pixels due to the imaging performance of the telescope and CCD.
Johnson $B$, Cousins $R_{\rm C}$, and Sloan Digital Sky Survey (SDSS) $z'$ filters were used for these observations.
Flat fielding observations were performed using the 0.6 m integrating sphere constructed in the same way as described by \citet{Sogame2005}.

Fifteen nights of photometric observations were obtained using the Ritchey--Chr\'etien telescope with an aperture of 0.36 m and a focal ratio of 8.0 at the Miyasaka Observatory.
This telescope was equipped with an SBIG STL-1001E CCD camera with a Kodak KAF-1001E detector whose format is 1024 $\times$ 1024 pixels.
This system produced image dimensions of \timeform{1".7}/pixel, yielding a field of view of \timeform{29'.4} $\times$ \timeform{29'.4}. 
The typical full-width half maximum of stellar images at this site was 3--5\timeform{"}.
Johnson $B$, Cousins $R_{\rm C}$, and SDSS $z'$ filters were used for the observations.
Dome flat fielding images were obtained from this 0.36 m telescope.

Seven nights of photometric observations were conducted using a refracting telescope with an aperture of 0.076 m and a focal ratio of 8.0 at the Nishiharima Astronomical Observatory.
This telescope is usually equipped with a 0.60 m reflecting telescope as a guide scope.
The data were recorded with an SBIG ST-9 CCD camera and Kodak KAF-0261 detector (512 $\times$ 512 pixels; angular resolution of \timeform{6".9}/pixel); sky field of \timeform{58'.8} $\times$ \timeform{58'.8}). 
The full-width half maximum of stellar images at this site was typically 1--3\timeform{"}.
The resolution is less than a pixel size, but star images spread out over several pixels due to the imaging performance of the telescope and CCD.
This telescope was equipped with Johnson $B$, Cousins $R_{\rm C}$, and SDSS $z'$ filters.
Measurements of flat fielding were made through the 0.6 m integrating sphere, which are similar to those made at the Sagamihara site.

A single night of photometric observation was recorded using a Zeiss reflecting telescope with an aperture of 0.60 m and a focal ratio of 12.5 at the Maidanak Astronomical Observatory. 
This telescope is equipped with a FLI 1000IMG CCD camera with a IMG1001E Kodak chip, which yielded an image format of 1024 $\times$ 1024 pixels and the projected area was \timeform{11'.7} $\times$ \timeform{11'.7}, corresponding to an angular resolution of \timeform{0".7}/pixel.
The typical full-width half maximum of stellar images at this site was ca. 1\timeform{"}.
Johnson $B$ and Cousins $R_{\rm C}$ filters were used for these observations.
Flat fielding images were taken during the evening and/or morning twilight using this 0.60 m telescope.

HD 268518 was selected for calibration in this study because of its close proximity to Vesta from 28 December 2005 to 8 January 2006, and because of its G0 spectral classification, which suggests that this star is a solar analog.
HD 268518 ($B = 8.193$ and $R_{\rm C}$ = 7.227 mag) is a standard star in the Johnson--Cousins photometric system, which is based on the Vega magnitude system \citep{Oja1996}.
Since HD 268518 is not a standard star in the SDSS photometric system, which is based on the AB magnitude system, it was calibrated with respect to the SDSS standard stars SA 97 249, SA 97 284, SA 97 288, SA 97 345, SA 97 351, SA 100 280, and SA 100 394 \citep{Smith2002}, resulting in $z' = 7.260$ mag.
The accuracy of the absolute photometry is estimated to be 0.01--0.04 mag.

All photometry was performed differentially relative to on-chip comparison stars, and then scaled to match the calibrated data. 
In this study, integrations of the standard star and all comparison stars without variable stars were rapidly alternated as both fields overlapped through the same airmass using the 0.064 m telescope with a very large field of view.
This observation method, which is made differentially with respect to nearby calibrated comparison stars, is considerably more efficient and accurate than only using catalog standards \citep{Harris1989}.
This approach also minimizes the effects of atmospheric extinction and weather changes.
The observational data, except for those obtained at the Sagamihara site, were calibrated by selecting comparison stars from the standard comparison stars, which were scaled to the photometric levels of HD 268518 as observed at the Sagamihara site.

Sidereal tracking was used for all photometric observations.
Since the integration times were short, the non-sidereal movement of Vesta during the integrations was smaller than the seeing size.
The dark images for correction were constructed from a median combination of 10--20 dark frames. 
Stacked flat fielding images for correction were created by a median combination of 10--20 flat fielding frames. 
All photometric image frames for an individual night were bias-subtracted and flat-fielded using Image Reduction and Analysis Facility (IRAF) software.
The fluxes of the asteroids and comparison stars were measured through circular apertures with a diameter of more than four times that of the full-width at half maximum size, using the APPOT task in IRAF.

The observed magnitude, $m_{\lambda}$(1, 1, $\alpha$), which is defined at a location of 1 AU from both the Sun and the Earth, and at a phase angle $\alpha\timeform{D}$ in a given filter, is calculated using the following equation:

\begin{eqnarray}
m_{\lambda}(1, 1, \alpha) =  m_{\lambda}(R_{\rm h}, \Delta, \alpha) - 5 \log(R_{\rm h} \Delta),
\label{eq:reduced magnitudes}
\end{eqnarray}

\noindent
where $m_{\lambda}(R_{\rm h}, \Delta, \alpha)$ is the observed apparent magnitude in the observed filter, $R_{\rm h}$ is the heliocentric distance, $\Delta$ is the geocentric distance, and $\alpha$ is the phase angle of Vesta.

\subsection{Spectroscopic observations}
\label{sec:Spectroscopic observations}
The spectroscopic observations were carried out at the Nishiharima Astronomical Observatory in Japan using the Nayuta 2.0 m telescope with an optical spectrograph and E2V CCD 42-40, yielding images of 2048 $\times$ 2048 pixels \citep{Hasegawa2006}.  
The spectrograph, the Medium And Low-dispersion Long-slit Spectrograph (MALLS), was attached to the f/12 Nasmyth focus of the Nayuta telescope.
A 150 line/mm grating with a dispersion of 338 \AA/mm in the first order was used.
The back-illuminated CCD has square 13.5 $\mu$m pixels, giving a dispersion of ca. 2.5 \AA/pixels in the wavelength direction.  
The slit length is \timeform{5'.0} in the cross-wavelength direction and the covered spectral range is ca. 0.49--0.98 $\mu$m.

The spectroscopic data were taken by the Kyoto Okayama Optical Low dispersion Spectrograph (KOOLS) (\cite{Ohtani1998}, \cite{Ishigaki2004}) attached to the f/18 Nasmyth focus of the Okayama Astrophysical Observatory's 1.88 m telescope.
A grism with a 6563 \AA/mm blaze was used.
The SITe ST-002A CCD has square 15.0 $\mu$m pixels, giving a dispersion of ca. 3.8 \AA/pixels in the wavelength direction.
The slit length is \timeform{4'.4} in the cross-wavelength direction and the covered spectral range is ca. 0.49--0.94 $\mu$m.

Particular care was taken on the choice of the slit width, in order to mitigate the consequences of atmospheric differential refraction \citep{Filippenko1982}. 
This is an important problem in asteroidal photometric spectroscopy.
The possible loss of photons at both ends of the spectrum cloud can lead to erroneous classification of the asteroid spectral type and a false calculation of spectral slope. 
Ideally, the slit should be perpendicular to the horizon, but the direction of the slit was not be able to be controlled due to issues with the instrument rotator at the Nasmyth focus.
Therefore, a wide-width 8" and 6" slit and was used for MALLS and KOOLS observations, respectively.

Vesta was observed during three runs on 27 February and 2 March in 2006, and on 28 January 2014. 
The typical seeing sizes on the three nights were $\sim$2".
The objects were observed near the meridian with an airmass lower than 2.0. 
Integration times were determined according to the brightness of each object.
Although the Nayuta telescope can track solar system objects in non-sidereal tracking mode, the data were obtained in sidereal tracking mode using a manual offset.
The objects were located in the center of the slit using a slit-viewer CCD.
The Okayama telescope was used to track asteroids with non-sidereal tracking.

Wavelength calibration frames were taken regularly during the night with light from an iron--neon--argon hollow cathode lamp.
The production of reflectance spectra from the wavelength-calibrated spectra is achieved through the division of the spectrum of a G2V Sun-like star.
To do this, HD 60298 and G63-51 were observed in 2006 and 2014, respectively.
Observations of the standard star were coordinated so that the star was observed at an airmass similar to those of the asteroid (i.e., the airmass difference was less than 0.1 in each case).
Since the airmass difference was 0.2 for KOOLS observations, the slope of the spectrum was corrected using the extinction curve of the solar analogue.
Flat fields were taken with a halogen lamp.
Frames of flat fields were obtained each night and averaged to obtain a high signal-to-noise ratio.
The observational details for the spectroscopic observations are listed in Table \ref{tab:spectroscopic circumstances}.

\begin{longtable}{cccccccc}
  \caption{Nightly details of spectroscopic observations.}\label{tab:spectroscopic circumstances}
  \hline
   Date & Time (UT) & $R_{\rm h}$  & $\Delta$  & $\alpha$ & Telescope & Integration & Airmass
\\ 
   $\rm[year.month.day]$ & [hour.minute.second] &  [AU] &  [AU] & [$\timeform{D}$]&  & time [sec] & 
\\
\endfirsthead
  \hline
   Date & Time (UT) & $R_{\rm h}$ & $\Delta$ & $\alpha$ & Telescope & Integration & Airmass
\\
  \hline
\endhead
  \hline
\endfoot
  \hline
\multicolumn{1}{@{}l}{\rlap{\parbox[t]{1.0\textwidth}{\small
}}}
\endlastfoot
  \hline
2006.Feb.27&13:43:09 -- 14:01:42&2.504&1.873& 20.3&2.0 m -- Nishiharima&4 $\times$ 100&1.24\\
2006.Mar.02&13:40:32 -- 13:49:24&2.501&1.905& 20.9&2.0 m -- Nishiharima&4 $\times$ 100&1.25\\
2006.Mar.02&14:22:55 -- 14:32:12&2.501&1.906& 20.9&2.0 m -- Nishiharima&4 $\times$ 100&1.43\\
2006.Mar.02&14:51:53 -- 15:00:30&2.501&1.906& 20.9&2.0 m -- Nishiharima&4 $\times$ 100&1.61\\
2006.Mar.02&15:20:29 -- 15:32:05&2.501&1.906& 20.9&2.0 m -- Nishiharima&4 $\times$ 100&1.87\\
2014.Jan.28&19:37:39 -- 19:39:03&2.287&1.862& 24.9&1.88 m -- Okayama&2 $\times$ 10&1.27\\
2014.Jan.28&19:46:06 -- 19:47:45&2.287&1.862& 24.9&1.88 m -- Okayama&2 $\times$ 10&1.26\\
2014.Jan.28&19:57:41 -- 19:59:50&2.287&1.862& 24.9&1.88 m -- Okayama&2 $\times$ 10&1.25\\

\end{longtable}

All data reduction was performed using the software package IRAF.
Although no significant change in flat field frames were observed from one night to another, for each night a separate flat field was used.
The bias from the over-scan region of the spectral CCD image was subtracted from all spectra.
After subtraction of bias, each object frame was divided by a normalized bias to correct the flat fielding flame.
The sky background in each object's spectrum was then fitted individually at each wavelength and subtracted.
The two-dimensional spectrum was then collapsed to one dimension, given that all the observed targets were point sources.
Following this, the dispersion solution for each iron--neon--argon spectrum was determined.
Solar analog standard star was used to compute reflectivities for spectra from the target.  
The solar spectra were treated in the same way as the asteroid spectra.  
The spectra were binned with a standard 19 boxcars for MALLS observations and 13 boxcars for KOOLS observations, and the spectral resolution was 0.05 $\mu$m.

\section{Photometric data}
\label{sec:Photometric data}
\subsection{Lightcurve correction}
\label{sec:Lightcurve correction}
\begin{figure}
  \begin{center}
    \FigureFile(80mm,80mm){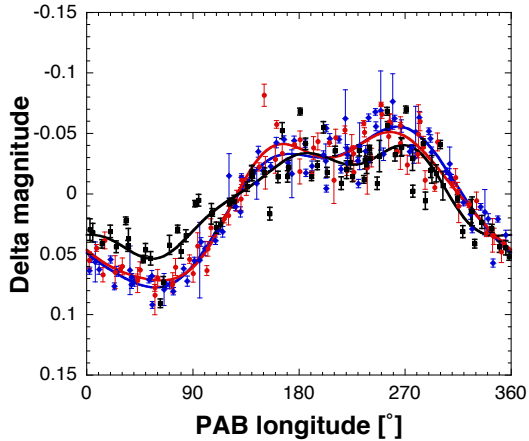}
  \end{center}
  \caption{Composite rotational lightcurve of Vesta.
$B$, $R_{\rm C}$, and $z'$ lightcurves of Vesta plotted as a function of the PAB longitude. 
PAB is the mean of the geocentric and heliocentric positions of the asteroid. 
The blue diamonds, red circles, and black squares indicate photometric data in the $B$, $R_{\rm C}$, and $z'$ band filters, respectively.
The solid line is a fourth-order fit obtained from Fourier analysis.
}
\label{fig:lightcurve}
\end{figure}

To obtain the mean magnitudes at the phase angle of 0$\timeform{D}$ for an accurate photometric function, the influence of rotational variation of Vesta must be eliminated.
The lightcurve of Vesta is already precisely known (e.g., \cite{Stephenson1951}).
For the purpose of the rotational variation correction, the lightcurves in the $B$, $R_{\rm C}$, and $z'$ bands were made as a function of the phase angle bisector (PAB) of Vesta longitude.
A coordinate system for Vesta has been defined by HST \citep{Thomas1997}.
Vesta's PAB longitudes were calculated from the mean values of the apparent planetographic longitude of Vesta and the apparent subsolar planetographic longitude of the Sun. 
The planetographic longitudes were obtained from the JPL Horizons ephemeris website.\footnotemark[1]
Composite lightcurves (plotted in Fig. \ref{fig:lightcurve}) contain data from multiple rotations with phase angles greater than 9$\timeform{D}$, where the phase angle changes by less than 0.1$\timeform{D}$ from one rotation to the next, because the change of phase angle is much faster at smaller phase angles.

The line present in all the lightcurves corresponds to a fourth-order Fourier series that best fits each respective data set. 
The lightcurves in the $B$, $R_{\rm C}$, and $z'$ bands are consistent with previous studies (\cite{Reynoldson1993}, \cite{Jaumann1996}, \cite{Nonaka2003}, \cite{Fulvio2008}). 
Based on these curves as a function of PAB longitude in each band, the corrections for rotational variation were applied to all photometric data obtained by ground-based observations.

\subsection{Combining the data with spacecraft observations}
\label{sec:Combination with data with spacecrafts}
It is difficult to observe Vesta from Earth at a phase angle more than $\sim$25$\timeform{D}$ due to its orbit.
The Rosetta spacecraft \citep{Glassmeier2007} made photometric observations at a phase angle of 52$\timeform{D}$ on the way to comet 67P/Churyumov--Gerasimenko on May 1 2010 \citep{Fornasier2011}. 
Between May 3 2011 and August 11 2012, Dawn also obtained photometric data at phase angles from 8$\timeform{D}$ to 108$\timeform{D}$ \citep{Li2013}.
These data from the Rosetta and Dawn spacecrafts are not able to be obtained from ground-based observations and, therefore, these data were used to supplement the photometric data from this study.

The Rosetta spacecraft conducted observations of Vesta using both narrow- and wide-angle cameras in the OSIRIS camera \citep{Keller2007} with red filters.
Absolute photometric values were obtained with an Bessel $R$ filter.
These values can be used without correction as the absolute values in the Cousins $R_{\rm C}$ band. 
Spectrophotometric data covering the wavelength range 0.27--0.99 $\mu$m were also taken by the OSIRIS camera at a phase angle of 52$\timeform{D}$.
The photometric values in the $B$ and $z'$ bands were estimated in combination with the $R_{\rm C}$ band photometric data, solar color, and reflectance data.
($B$ $\mathrm{-}$ $R_{\rm C}\mathrm{)_{\odot}}$ was taken from \citet{Ramirez2012}.
Based on the transformation equations between the SDSS and Johnson--Cousins photometric systems \citep{Rodgers2006} and the solar color from SDSS \citep{Ivezic2001}, ($R_{\rm C}$ $\mathrm{-}$ $z'\mathrm{)_{\odot}}$ was determined.
The $B$ and $z'$ band reflectances for Vesta were acquired by interpolating spectrophotometric data from Rosetta \citep{Fornasier2011}.

The Dawn spacecraft collected data for a disk-integrated phase function in visible wavelengths with a Framing Camera \citep{Schroder2013} through a clear filter.
The effective wavelength of the clear filter is 0.732 $\mu$m, which is similar to the $R_{\rm C}$ band.
Given that the filter band of photometric values in \citet{Li2013} reported the Johnson $V$ band, the data were converted through the V band to the $R_{\rm C}$ band using the solar color ($V$ $\mathrm{-}$ $R_{\rm C}\mathrm{)_{\odot}}$ and spectrum of Vesta.

\begin{figure}
  \begin{center}
    \FigureFile(80mm,80mm){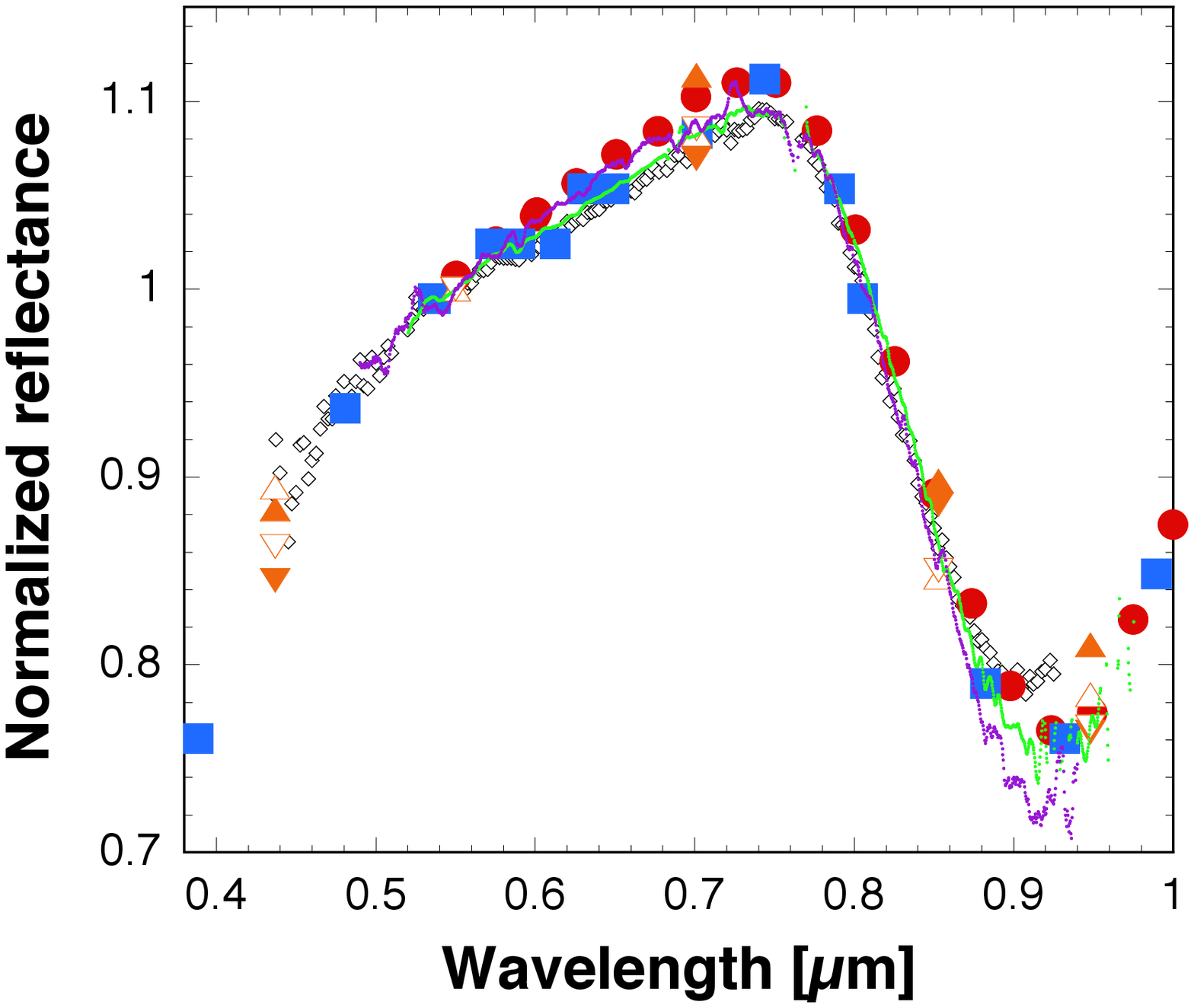}
  \end{center}
  \caption{Comparison of visible spectra for Vesta.
Orange triangles, red circles, black diamonds, blue squares, green dots, and purple dots indicate spectra from \citet{Zellner1985}, \citet{Hicks1998}, Xu et al. (1995), \citet{Bus2002}, \citet {Fornasier2011}, this study (Nishiharima), and this study (Okayama), respectively.
The spectra derived from the Nishiharima and Okayama observations are the combination of four spectra from 2 March 2006 and three spectra from 28 January 2006, respectively.
Spectroscopic data for open inverted triangles, open diamonds, open triangles, filled inverted triangles, filled triangles, filled circles, green dots, purple dots, and filled squares are obtained at phase angles of 7.9, 11.9, 15.6, 16.0, 17.5, 19.0, 20.9, 24.9, and 52.5$\timeform{D}$, respectively.
}
\label{fig:spectrum}
\end{figure}

Fig. \ref{fig:spectrum} shows the spectra of Vesta acquired with the Nayuta and Okayama telescopes in this study, as compared with previous studies.
The combined spectra from this study is in good agreement with previous studies.
For adjustment to the $R_{\rm C}$ band, the spectrum of Vesta obtained in this study was used.
The solar color is similar to the value reported by \citet{Ramirez2012}.
Using these values, the conversion to the $R_{\rm C}$ band was carried out.

\section{Phase function}
\label{sec:Phase function}
The combined $B$, $R_{\rm C}$, and $z'$ band data from this study, along with those from \citet{Fornasier2011} and \citet{Li2013}, were used for the photometric phase function.
The values obtained at each phase angle were compiled in steps of 0.1$\timeform{D}$.
To know the properties of the opposition effect and phase function, the data were fitted with three function models (i.e., the IAU $H$--$G$ phase function model, the Shevchenko function model, and the Hapke model).

A routine was employed to obtain a solution that minimizes the $\chi^2$ values for all models, and in which all parameters can be varied to achieve this.
The curve fit is based on the Levenberg--Marquardt algorithm, which is commonly employed to solve nonlinear least-squares problems.
The $\chi^2$ values are defined as:

\begin{eqnarray}
\chi^2 = \frac{1}{N} \sum^{N}_{n=1}\left[M_{\lambda}(1, 1, \alpha) -m_{\lambda}(1, 1, \alpha)\right]^2,
\label{eq:least-squares}
\end{eqnarray}

\noindent
where $N$ is the number of data points, and $M_{\lambda}(1, 1, \alpha)$ is the model magnitude at 1 AU from both the Sun and the Earth and at a phase angle $\alpha\timeform{D}$ in a given filter. 
Upon $\chi^2$ minimization, the corresponding parameter value was included in the final model set of parameters.
The error in the parameter value is the standard error.

\subsection{IAU H--G phase function}
\label{sec:IAU H--G phase function}
The IAU $H$--$G$ phase function model is a simple two-parameter empirical function that was accepted by Commission 20 of the International Astronomical Union \citep{Bowell1989}.
This is the most common phase function of asteroids and has the following form:

\begin{eqnarray}
M_{\lambda}(1, 1, \alpha) = H_{\lambda} - 2.5 \log \left[(1 - G_{\lambda}) \Phi_1(\alpha) + G_{\lambda} \Phi_2(\alpha)\right], 
\label{eq:HG-eq}
\end{eqnarray}

\noindent 
where $H_{\lambda}$ is the absolute magnitude in the observed band at a phase angle of 0$\timeform{D}$, and $G_{\lambda}$ is the so-called slope parameter, which describes the shape of the magnitude phase function. 
$\Phi_1(\alpha)$ and $\Phi_2(\alpha)$ are two basis functions normalized at unity for $\alpha$ = 0$\timeform{D}$. 
These functions are accurately approximated by:

\begin{eqnarray}
\begin{array}{lcl}
\Phi_1(\alpha) = \exp \left(-3.33 \tan^{0.63}\frac{\alpha}{2}\right),  \\
\Phi_2(\alpha) = \exp \left(-1.87 \tan^{1.22}\frac{\alpha}{2}\right).  \\
\end{array}
\label{eq:HG-phi}
\end{eqnarray}

Table \ref{tab:slope parameter} lists the $H_{\lambda}$ and $G_{\lambda}$ values obtained for Vesta in this and previous studies.
$H_{R_{\rm C}}$ and $G_{R_{\rm C}}$ were fitted by changing the adaptation range of the phase angle to determine a phase angle range consistent with the IAU $H$--$G$ phase function (see sans-serif fonts in Table \ref{tab:slope parameter}).
Taking a phase angle range provided from the Earth into consideration, $H_{R_{\rm C}}$ and $G_{R_{\rm C}}$ decrease whilst the adopted minimum phase angle decreases due to the opposition effect.
If the minimum phase angle is more than 7$\timeform{D}$, which is the onset point of the opposition effect of asteroids \citep{Scaltriti1980}, then the obtained solution is scattered and lost.
$H_{R_{\rm C}}$ and $G_{R_{\rm C}}$ do not change with an increase in the adopted largest angle until 81.3$\timeform{D}$ due to the linear nature of the magnitude diminution.
When the largest phase angle is found at more than 81.3$\timeform{D}$, the values cannot be correct.
Therefore, the appropriate phase angle range for determination of $H_{R_{\rm C}}$ and $G_{\rm R_{c}}$ is 0.1--81.3$\timeform{D}$ (see bold fonts in Table \ref{tab:slope parameter}).

\begin{longtable}{ccccl}
  \caption{Comparison of IAU $H$--$G$ phase functions for Vesta.}\label{tab:slope parameter}\label{tab:slope parameter}
  \hline
   $H_{\lambda}$ & $G_{\lambda}$ & Filter band & $\alpha$ &References 
\\ 
   $\rm[mag]$ &  & ([$\mu$m])\footnotemark[$*$] & [$\timeform{D}$] & 
\\
\endfirsthead
  \hline
   $H_{\lambda}$ & $G_{\lambda}$ & Filter band & $\alpha$ &References 
\\
  \hline
\endhead
  \hline
\endfoot
  \hline
\multicolumn{1}{@{}l}{\rlap{\parbox[t]{1.0\textwidth}{\small
\footnotemark[$*$]Values in parentheses are the effective wavelength of the Framing Camera onboard the Dawn spacecraft.\\
\footnotemark[$\dagger$]Data in \citet{Fornasier2011} include preliminary data from this study \citep{Hasegawa2009}.\\
\footnotemark[$\ddagger$]Fitting errors appear within brackets below each parameter value. \\
}}}
\endlastfoot
  \hline
\bf{3.96}&\bf{0.30}&\boldmath $B$&\bf{0.1 -- 52.3}&\bf{This study}\\
$\langle0.01\rangle$&$\langle0.02\rangle$&&&Fitting error\footnotemark[$\ddagger$]\\
-----&0.32&F8 (0.44) &3.8 -- 25.7& \citet{Reddy2012}\\
  \hline
\bf{2.82}&\bf{0.29}&\boldmath $R_{\rm C}$&\bf{0.1 -- 81.3}&\bf{This study}\\
$\langle0.01\rangle$&$\langle0.01\rangle$&&&Fitting error\footnotemark[$\ddagger$]\\
\sf{2.81}&\sf{0.27}&$\sf{\mathit{R_{C}}}$&\sf{0.1 -- 23.8}&\sf{This study}\\
\sf{2.85}&\sf{0.31}&$\sf{\mathit{R_{C}}}$&\sf{1.0 -- 23.8}&\sf{This study}\\
\sf{2.81}&\sf{0.27}&$\sf{\mathit{R_{C}}}$&\sf{7.0 -- 23.8}&\sf{This study}\\
\sf{2.82}&\sf{0.28}&$\sf{\mathit{R_{C}}}$&\sf{0.1 -- 43.1}&\sf{This study}\\
\sf{2.82}&\sf{0.28}&$\sf{\mathit{R_{C}}}$&\sf{0.1 -- 52.3}&\sf{This study}\\
\sf{2.87}&\sf{0.39}&$\sf{\mathit{R_{C}}}$&\sf{0.1 -- 108.6}&\sf{This study}\\
2.80&0.27&$R_{\rm C}$&0.1\footnotemark[$\dagger$] -- 52.3& \citet{Fornasier2011}\\
-----&0.29&F7 (0.65) &3.8 -- 25.7& \citet{Reddy2012}\\
  \hline
\bf{3.08}&\bf{0.25}&\boldmath $z'$&\bf{0.1 -- 52.3}&\bf{This study}\\
$\langle0.01\rangle$&$\langle0.01\rangle$&&&Fitting error\footnotemark[$\ddagger$]\\
-----&0.36&F4 (0.92) &3.8 -- 25.7& \citet{Reddy2012}\\
  \hline
3.16&0.34&$V$&-----&Minor Planet Circ. 11095 (1986)\\
3.38&0.47&$V$&5.9 -- 25.5& \citet{Lagerkvist1987}\\
3.28&0.41&$V$&4.9 -- 23.4& \citet{Lagerkvist1988}\\
3.25&0.35&$V$&1.4 -- 22.6& \citet{Lagerkvist1989}\\
3.40&0.33&$V$&1.7 -- 23.7& \citet{Lagerkvist1990}\\
3.20&0.32&$V$&-----&Minor Planet Circ. 17256 (1990)\\
3.32&0.42&$V$&1.3 -- 28.1& \citet{Rock1990}\\
-----&0.35&$V$&1.2 -- 26.1& \citet{Lagerkvist1990}\\
3.19&0.32&$V$&4.9 -- 25.3& \citet{Piironen1997}\\
3.14&0.32&$V$&0.4 -- 24.8& \citet{Velichko2008}\\
-----&0.23&F2 (0.56) &3.8 -- 25.7& \citet{Reddy2012}\\
3.2&0.28&$V$&1.7 -- 108.6& \citet{Li2013}\\

\end{longtable}

\begin{figure}
  \begin{center}
    \FigureFile(80mm,80mm){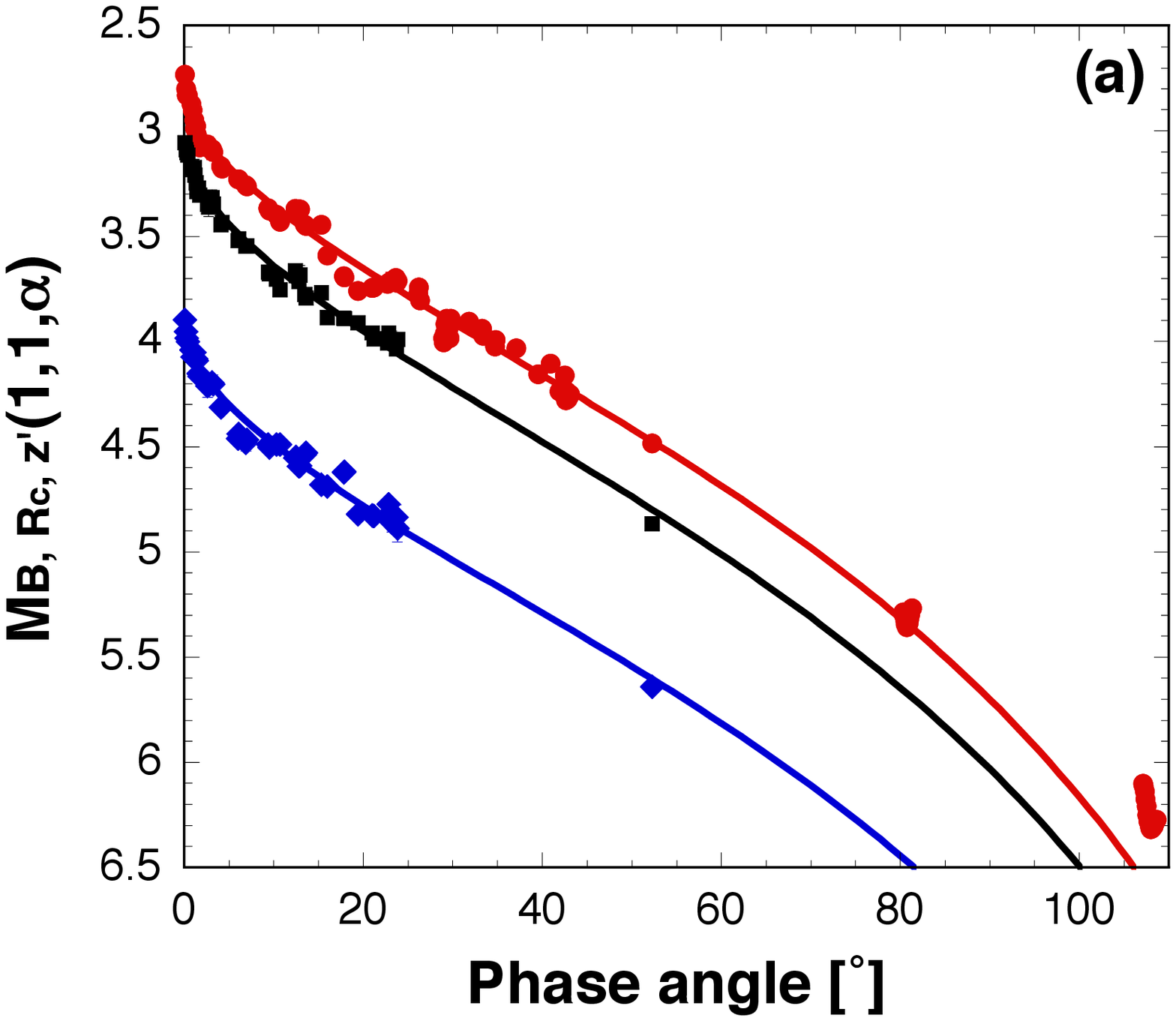}
    \FigureFile(80mm,80mm){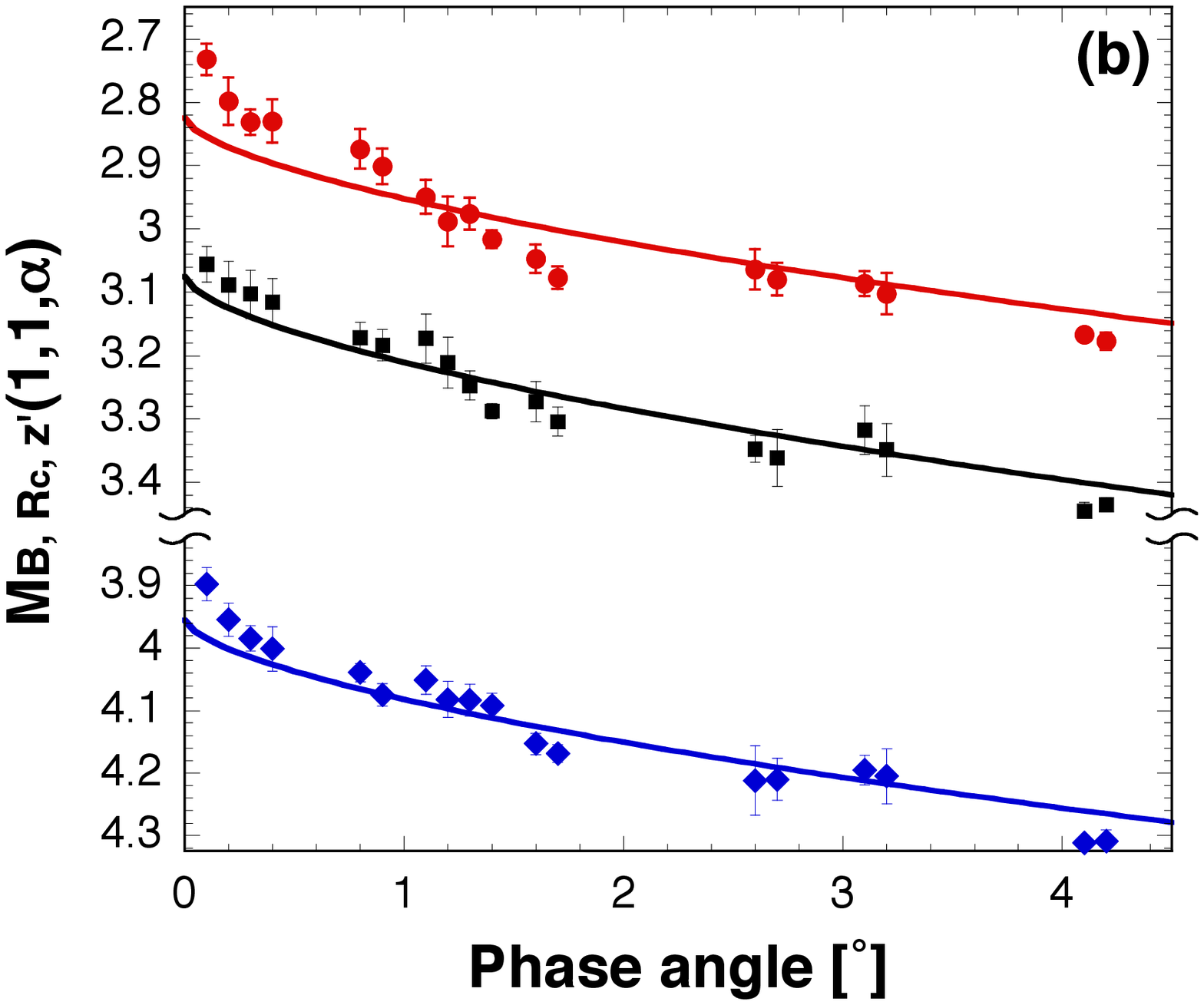}
  \end{center}
  \caption{Fitted phase curves for Vesta using the IAU $H$--$G$ phase function. (a) and (b) show the whole curves and expanded views at small phase angles, respectively.
The blue diamonds, red circles, and black squares are photometric data from the $B$, $R_{\rm C}$, and $z'$ band filters, respectively.
The lines show the best-fit phase curves obtained with the IAU $H$--$G$ phase function.
}
\label{fig:hg}
\end{figure}

The $H_{\lambda}$ and $G_{\lambda}$ values that most closely fit the data give $H_{B}$ = 3.96 $\pm$ 0.01, $G_{B}$ = 0.30 $\pm$ 0.02, $H_{R_{\rm C}}$ = 2.82 $\pm$ 0.01, $G_{R_{\rm C}}$ = 0.29 $\pm$ 0.01, $H_{z'}$ = 3.08 $\pm$ 0.02, and $G_{z'}$ = 0.25 $\pm$ 0.01 mag.
The phase curves for Vesta in the $B$, $R_{\rm C}$, and $z'$ bands using the IAU $H$--$G$ phase function are shown in Fig. \ref{fig:hg}.
The values from the $B$, $R_{\rm C}$, and $z'$ bands from this study are consistent with those of \citet{Fornasier2011} and \citet{Reddy2012}.
Although the wavelengths used differ, $G_{\lambda}$ values from this study are similar to those of previous studies (Minor Planet Circ. 17256, \cite{Lagerkvist1990}, \cite{Piironen1997}, \cite{Velichko2008}, \cite{Li2013}).

\subsection{Shevchenko function}
\label{sec:Shevchenko function}
The Shevchenko phase function model is a simple three-parameter empirical function that was proposed by \citet{Shevchenko1996}.
This function is a direct definition of the amplitude of the opposition effect and has the following form:

\begin{eqnarray}
M_{\lambda}(1, 1, \alpha) = C_{\lambda} - \frac{a_{\lambda}}{1 +\alpha} + b_{\lambda} \alpha ,
\label{eq:Shevchenko-eq}
\end{eqnarray}

\noindent 
where $a_{\lambda}$ is a parameter to characterize the opposition effect amplitude and $b_{\lambda}$ is the parameter describing the linear part of the magnitude phase dependence. 
$C_{\lambda}$ is a constant defined by $C_{\lambda}$ = $m_{\lambda}$(1, 1, 0) + $a_{\lambda}$. 

To determine the phase angle range for the Shevchenko phase function, the parameters $a_{R_{\rm C}}$ and $b_{R_{\rm C}}$ were fitted by changing the adaptation range of the phase angle. 
The precision of $a_{R_{\rm C}}$ increases when the angle is reduced.
When the minimum phase angle is more than 7$\timeform{D}$, the obtained $a_{R_{\rm C}}$ cannot be correct.
This behavior is the same as when fitting results with the IAU $H$--$G$ phase function (Section \ref{sec:IAU H--G phase function}).
The parameter $b_{R_{\rm C}}$ is changed until 81.3$\timeform{D}$, and such that the linear fit degrades.
Therefore, the phase angle range for assignment of the parameters in the $R_{\rm C}$ band is changed from 0.1$\timeform{D}$ to 81.3$\timeform{D}$.
The parameters $a_{\lambda}$ and $b_{\lambda}$ obtained for Vesta in this study are listed in Table \ref{tab:Shevchenko function}.

\begin{figure}
  \begin{center}
    \FigureFile(80mm,80mm){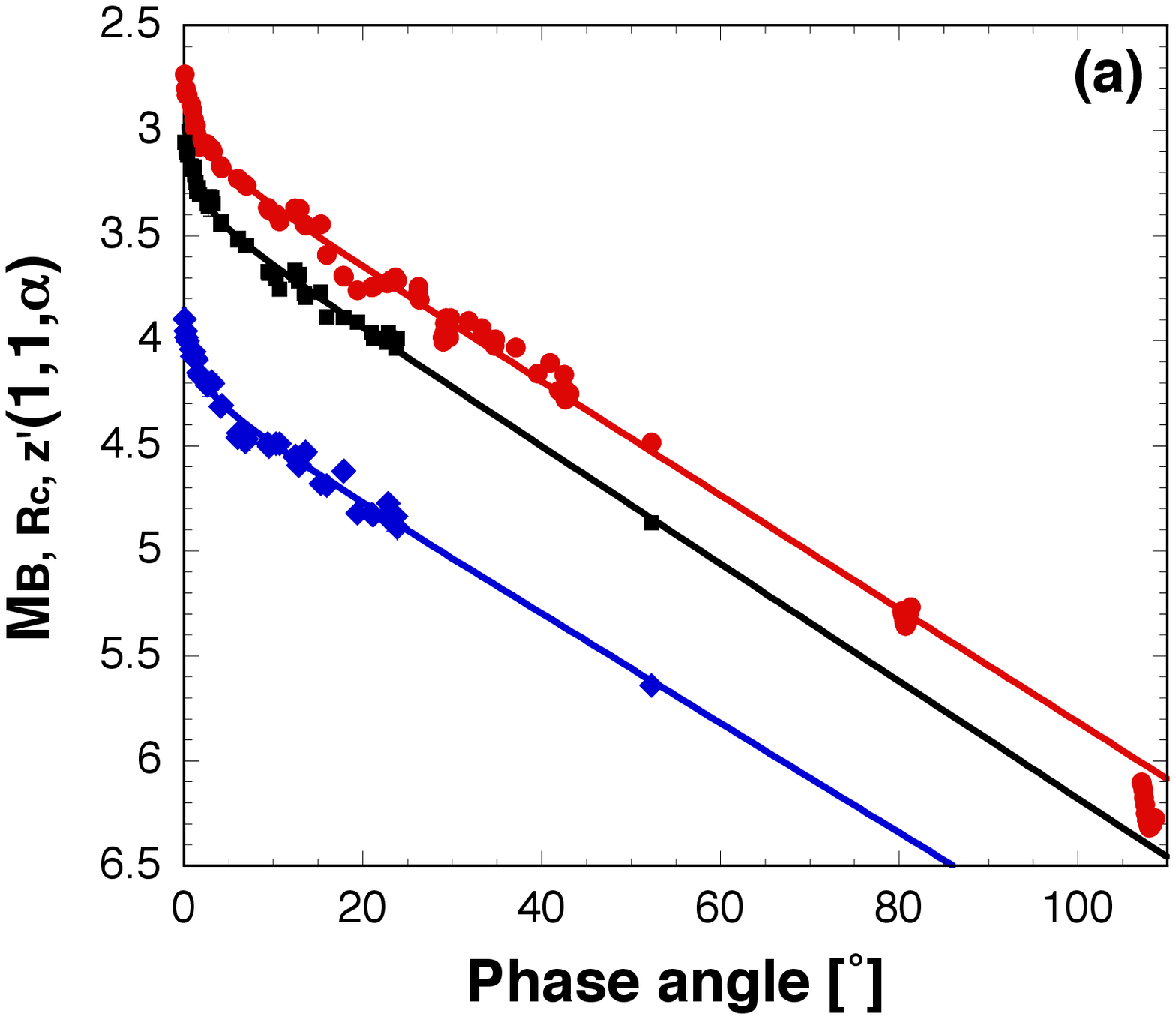}
    \FigureFile(80mm,80mm){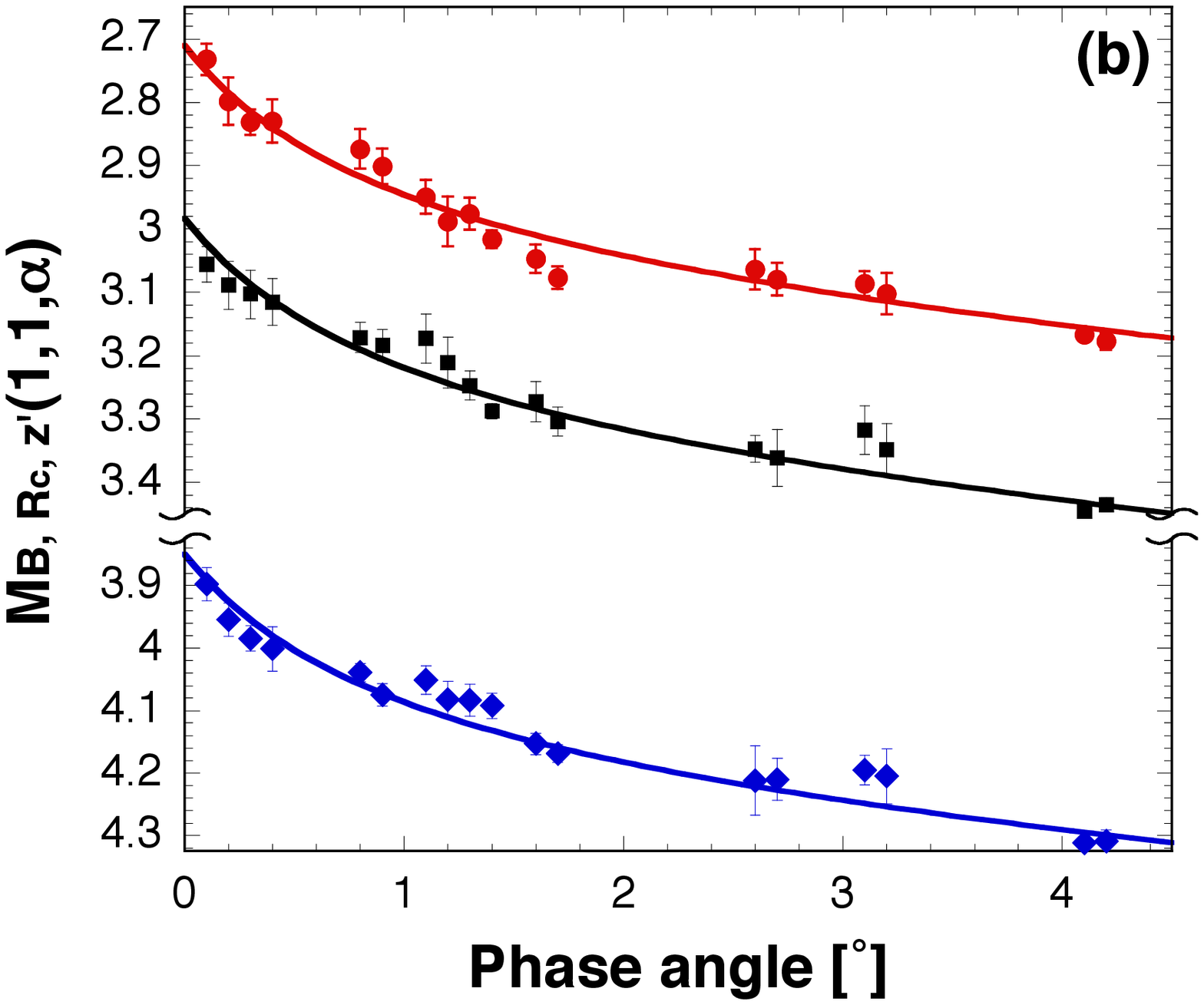}
  \end{center}
  \caption{Fitted phase curves for Vesta using the Shevchenko function. 
(a) and (b) show whole curves and expanded views at small phase angles, respectively.
The blue diamonds, red circles, and black squares are photometric data in the $B$, $R_{\rm C}$, and $z'$ band filters, respectively.
The lines show the best-fit phase curves obtained with the Shevchenko function.
}
\label{fig:Shevchenko}
\end{figure}

The $M_{\lambda}$(1, 1, 0) value for Vesta that fits the data most closely based on the Shevchenko function yields $M_{B}$(1, 1, 0) = 3.85 $\pm$ 0.01, $M_{R_{\rm C}}$(1, 1, 0) = 2.71 $\pm$ 0.01, and $M_{z'}$(1, 1, 0) = 2.98 $\pm$ 0.01 mag.
The phase curves for Vesta in the $B$, $R_{\rm C}$, and $z'$ bands using the Shevchenko function are shown in Fig. \ref{fig:Shevchenko}.

\begin{longtable}{ccccccl}
  \caption{Comparison of the Shevchenko function for Vesta.}\label{tab:Shevchenko function}
  \hline
   $M_{\lambda}$(1, 1, 0) & $a_{\lambda}$ & $b_{\lambda}$ &$C_{\lambda}$ &Filter band & $\alpha$ &References
\\ 
   $\rm[mag]$ &  & [mag/$\timeform{D}$] & $\rm[mag]$ &(eff. $\lambda$ [$\mu$m]) & [$\timeform{D}$] &
\\
\endfirsthead
  \hline
   $M_{\lambda}$(1, 1, 0) & $a_{\lambda}$ & $b_{\lambda}$ &$C_{\lambda}$ &Filter band & $\alpha$ &References
\\
  \hline
\endhead
  \hline
\endfoot
  \hline
\multicolumn{1}{@{}l}{\rlap{\parbox[t]{1.0\textwidth}{\small
\footnotemark[$*$]Fitting errors appear within brackets below each parameter value. \\
}}}
\endlastfoot
  \hline
\bf{3.85}&\bf{0.427}&\bf{0.0259}&\bf{4.271}&\boldmath $B$&\bf{0.1 -- 52.3}&\bf{This study}\\
$\langle0.01\rangle$&$\langle0.037\rangle$&$\langle0.0008\rangle$&$\langle0.018\rangle$&&&Fitting error\footnotemark[$*$]\\
  \hline
\bf{2.71}&\bf{0.416}&\bf{0.0269}&\bf{3.127}&\boldmath $R_{\rm C}$&\bf{0.1 -- 81.3}&\bf{This study}\\
$\langle0.01\rangle$&$\langle0.037\rangle$&$\langle0.0003\rangle$&$\langle0.012\rangle$&&&Fitting error\footnotemark[$*$]\\
  \hline
\bf{2.98}&\bf{0.416}&\bf{0.0278}&\bf{3.400}&\boldmath $z'$&\bf{0.1 -- 52.3}&\bf{This study}\\
$\langle0.01\rangle$&$\langle0.031\rangle$&$\langle0.0008\rangle$&$\langle0.014\rangle$&&&Fitting error\footnotemark[$*$]\\

\end{longtable}

\subsection{Hapke model}
\label{sec:Hapke model}
The Hapke model is a five-parameter, quasi-experimental function (\authorcite{Hapke1981} \yearcite{Hapke1981}, \yearcite{Hapke1984}, \yearcite{Hapke1986}).
This model is based on a physical description of the scattering behavior of a particulate surface and is, therefore, more easily interpretable than empirical models, such as the IAU $H$--$G$ phase and Shevchenko functions.
Still, it is difficult to characterize the physical interpretations based on parameters of the Hapke model. 
There is no strong correlation between Hapke parameters and the actual physical particle properties (\authorcite{Shepard2007}  \yearcite{Shepard2007}, \yearcite{Shepard2011}).
\citet{Hapke2002} revised this model to incorporate an opposition surge element that takes into account coherent backscatter, and also included an additional term for multiple scattering. However, this model is typically not used unless the phase angle is low (less than 2$\timeform{D}$).
Given that our photometric observations include low phase angles up to 0.1$\timeform{D}$, the updated Hapke mode (a seven-parameter function), as well as the nominal Hapke model, was used in this study.

\citet{Hapke2008} developed the model to express a porosity dependence.
However, the updated Hapke model \citep{Hapke2008} was not used in this study for two reasons, explained below.
First, \citet{Li2013} attempted to fit their data using the Hapke model with the relevant porosity parameter \citep{Hapke2008}, but they could not obtain any meaningful results for the porosity parameter. 
Data for this study contain disk-integrated photometric measurements including small phase angles (less than 1$\timeform{D}$), but we do not have access to large quantities of disk-resolved photometric data such as \citet{Li2013}. 
It can be predicted easily that the new Hapke model will not yield any meaningful porosity parameter using data from this study. 
Second, it is hard to obtain a unique solution when the number of unknowns is increased.
By reducing the number of fitting parameters by one (in this case, the porosity parameter), a stable solution can be obtained.

The magnitudes in each band were converted to the logarithmic form of $I_{\lambda}/F_{\lambda}$ (where $\pi F_{\lambda}$ is the incident solar flux and $I_{\lambda}$ is the light scattered from the surface).

$I_{\lambda}/F_{\lambda}$ is expressed as follows: 

\begin{eqnarray}
-\frac{5}{2} \log \left(\frac{I_{\lambda}}{F_{\lambda}}\right) = M_{\lambda}(1, 1, \alpha) - M_{\lambda \odot} -\frac{5}{2} \log
\left( \frac{\pi}{A_{rea}}\right) + M_c,
\label{eq:hapke1}
\end{eqnarray}

\noindent
where $M_{\lambda \odot}$ is the magnitude of the Sun at 1 AU in the observed filter (the $B$-band absolute magnitude is taken from \cite {Allen1976}, the $R_{\rm C}$ and $z'$-band absolute magnitudes are obtained from a combination with $M_{B \odot}$ and the solar colors; \cite{Ramirez2012}, \cite{Rodgers2006}, \cite{Ivezic2001}), $A_{rea}$ is the geometrical cross-section of Vesta in m$^2$ \citep{Russell2012}, and $M_c$ = $-5\log (1.4960 \times 10^{11} [m])$ = $-55.87$ is a constant.
Given that Vesta is almost spherical \citep{Russell2012}, we used the Hapke model using an integral phase function for a spherical body with a rough surface.
The Hapke equation for a spherical body with a rough surface (\authorcite{Hapke1984} \yearcite{Hapke1984}, \yearcite{Hapke1986}) is given by:

\begin{eqnarray}
\nonumber
\frac{I_{\lambda}}{F_{\lambda}} =
 \left[\left(\frac{w_{\lambda}}{8}\left[\left(1+B_{SH_{\lambda}}(\alpha)\right)P_{\lambda}(\alpha)-1\right]+\frac{r_{0_{\lambda}}}{2}(1-r_{0_{\lambda}})\right) 
   \left(1-\sin\frac{\alpha}{2} \tan \frac{\alpha}{2}
   \ln \left[\cot\frac{\alpha}{4}\right]\right) \right.\\  
   \left.
   +\frac{2}{3}{r_{0_{\lambda}}}^2\left(\frac{\sin \alpha+(\pi-\alpha)\cos \alpha}{\pi}\right)\right]
   \left(1+B_{CB_{\lambda}}(\alpha)\right)K_{\lambda}(\alpha,\bar \theta_{\lambda}),~~
\label{eq:hapke2-main}
\end{eqnarray}

\noindent
where $w_{\lambda}$ is the single-particle scattering albedo in a particular band.
The term $r_{0_{\lambda}}$ is given by: 

\begin{eqnarray}
r_{0_{\lambda}} = \frac{1-\sqrt{1-w_{\lambda}}}{1+\sqrt{1-w_{\lambda}}},
\label{eq:hapke-r0}
\end{eqnarray}

\noindent
The one-term Henyey--Greenstein single-particle phase function solution \citep{Henyey1941} is:  

\begin{eqnarray}
P_{\lambda}(\alpha)=\frac{\left(1-{g_{\lambda}}^2\right)}{\left(1+2g_{\lambda} \cos (\alpha) + {g_{\lambda}}^2\right)^{3/2}}, 
\label{eq:hapke-OTHGF}
\end{eqnarray}

\noindent
where $g_{\lambda}$ is an asymmetry factor. 
$K_{\lambda}(\alpha,\bar \theta_{\lambda})$ corrects the surface roughness with the surface roughness parameter $\bar \theta_{\lambda}$ \citep{Hapke1984}.
The opposition effect term of the shadow-hiding ($B_{SH_{\lambda}}(\alpha)$) is given by:

\begin{eqnarray}
B_{SH_{\lambda}}(\alpha) = \frac{B_{S0_{\lambda}}}{1+\frac{1}{h_{S_{\lambda}}} \tan\frac{\alpha}{2}},
\label{eq:hapke-SWOE}
\end{eqnarray}

\noindent
where $B_{S0_{\lambda}}$ describes the amplitude of the shadow-hiding opposition effect and $h_{S_{\lambda}}$ is the width of the shadow-hiding opposition effect peak in radian. 
The coherent backscatter opposition effect ($B_{CB_{\lambda}}(\alpha)$) \citep{Hapke2002} is given by:

\begin{eqnarray}
B_{CB_{\lambda}}(\alpha) = B_{C0_{\lambda}} \frac{1 + \frac{1 - \exp \left({-\frac{1}{h_{c_{\lambda}}} \tan\frac{\alpha}{2}}\right)}{\frac{1}{h_{c_{\lambda}}} \tan\frac{\alpha}{2}}}{2\left(1 + \frac{1}{h_{c_{\lambda}}} \tan\frac{\alpha}{2}\right)^2},
\label{eq:hapke-CBOE}
\end{eqnarray}

\noindent
where $B_{C0_{\lambda}}$ describes the amplitude of the coherent backscatter opposition effect and $h_{C_{\lambda}}$ is the width of the coherent backscatter opposition effect peak in radian. 

Parameters in the Hapke model are constrained by the strength and shape of the phase curves:
$w_{\lambda}$ is determined by the absolute values of the phase curves; 
$g_{\lambda}$ is controlled by the shape of the curves at all phase angles; 
$\bar \theta_{\lambda}$ is affected by the shape of the phase function at large phase angles; 
$B_{S0_{\lambda}}$ and $h_{S_{\lambda}}$ are obtained from the shape and the intensity of the phase curve in the phase angle range between 0 and 20$\timeform{D}$; and 
$B_{C0_{\lambda}}$ and $h_{C_{\lambda}}$ are provided by the intensity of the spike phase curve in a phase angle range within several degrees.

Photometric data in $\mathrm{R_{C}}$ were covered over a phase angle from 0.1$\timeform{D}$ to 108.6$\timeform{D}$, but those in the B and z' bands were obtained from 0.1$\timeform{D}$ and 52.3$\timeform{D}$.
\citet{Helfenstein1989} indicated that $\bar \theta$ affected the phase function at phase angles greater than 40$\timeform{D}$.
Data in the $B$ and $z'$ bands at phase angles in excess of 40$\timeform{D}$ are indispensable to constrain $\bar \theta_B$ and $\bar \theta_{z'}$.
\citet{Li2013} showed that $\bar \theta$ in the visible wavelength range from 0.44 to 0.96 $\mu$m, in both cases, is concentrated within a small range of $\pm$ 2$\timeform{D}$ at ca. 18$\timeform{D}$. This study also identified any wavelength independence with respect to $\bar \theta$.
To reduce the parametric fit to the model, the $\bar \theta_{\lambda}$ value of \citet{Li2013} was adopted in our study.

The phase angle range in the $R_{\rm C}$ band for the IAU $H$--$G$ and Shevchenko functions was adopted from 0.1$\timeform{D}$ to 81.3$\timeform{D}$ (see Sections \ref{sec:IAU H--G phase function} and \ref{sec:Shevchenko function}).
For the Hapke model, the phase angle ranged between 0.1$\timeform{D}$ and 81.3$\timeform{D}$.

The geometric albedo ($A_{p_{\lambda}}$) is defined as the fraction of incident light scattered by the surface at a phase angle 0$\timeform{D}$ relative to a similar sized Lambert disk based on the same observing geometry. 
The bond albedo ($A_{B_{\lambda}}$) is defined as the total radiation reflected from an object as compared with the total incident radiation from the Sun, which is given by $A_{B_{\lambda}}$ = $q_{\lambda}$ $A_{p_{\lambda}}$, where $q_{\lambda}$ is the value of the phase integral defined as: 

\begin{eqnarray}
q_{\lambda} = 2 \int_0^{\pi} \frac{\phi_{\lambda}(\alpha)}{\phi_{\lambda}(0)}~\sin(\alpha)~d\alpha.
\end{eqnarray}

\noindent
where $\phi_{\lambda}(\alpha)$ is defined in Eq. \ref{eq:hapke1} as $I_{\lambda}/F_{\lambda}$.

Two independent determinations of the Hapke parameters were performed with and without inclusion of coherent backscatter opposition effect terms.
The Hapke parameters for Vesta, including a comparison with previous studies, are listed in Table \ref{tab:Hapke function}.

\begin{figure}
  \begin{center}
    \FigureFile(80mm,80mm){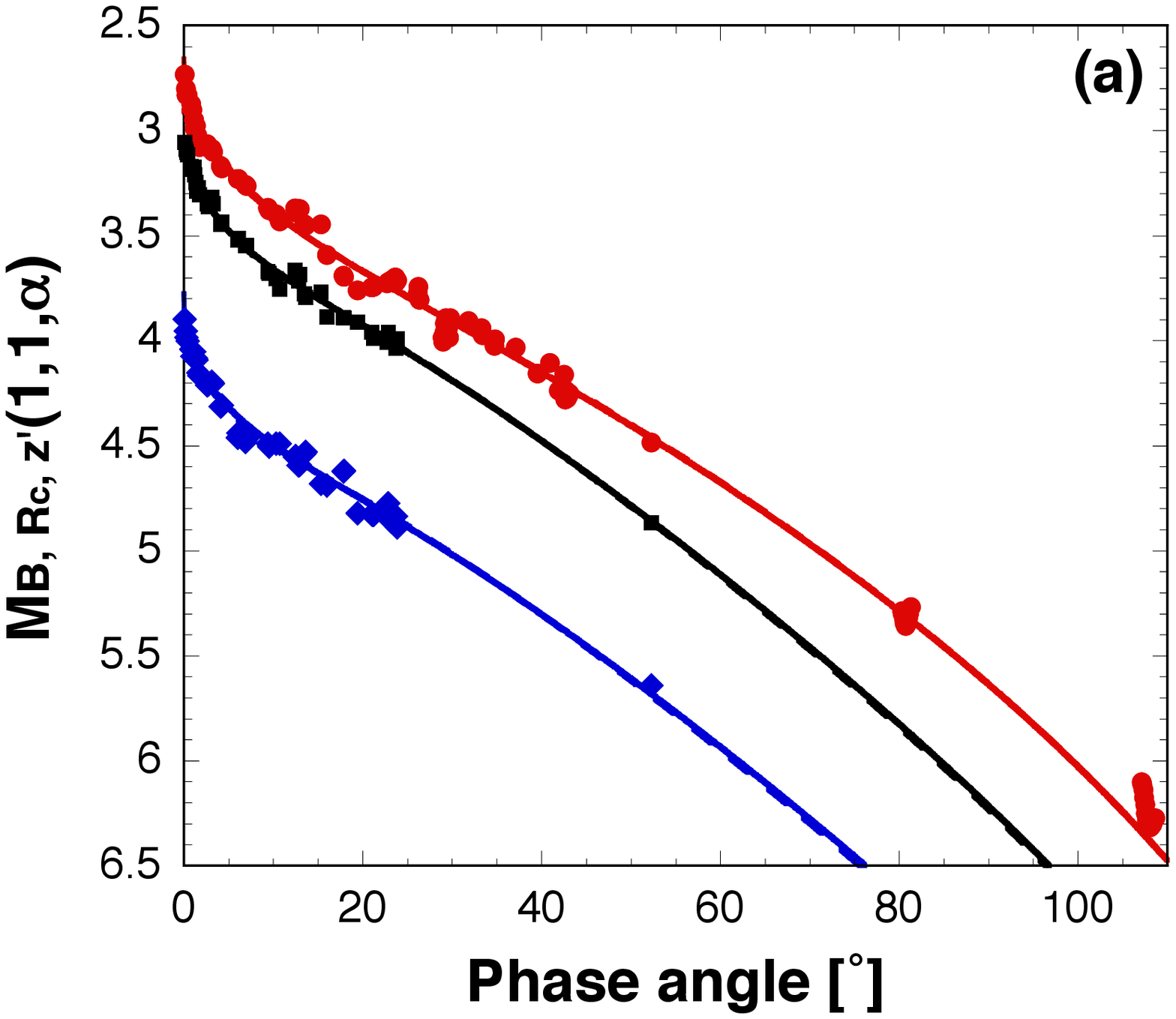}
    \FigureFile(80mm,80mm){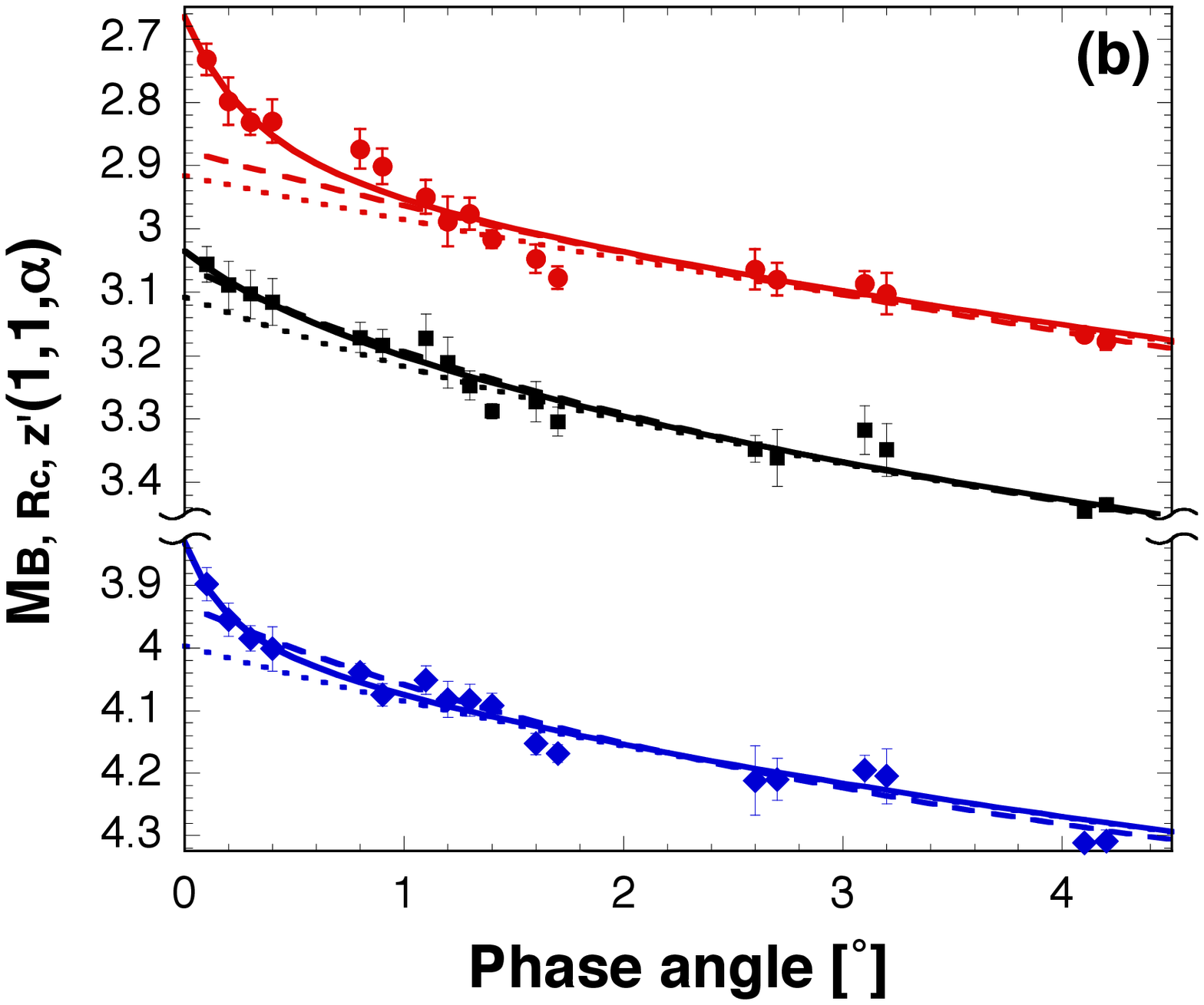}
  \end{center}
  \caption{Fitted phase curves for Vesta obtained with the Hapke model. 
(a) and (b) show whole curves and expanded views of the curves at small phase angles, respectively.
The blue diamonds, red circles, and black squares are photometric data in the $B$, $R_{\rm C}$, and $z'$ band filters, respectively.
The solid, dotted, and dashed lines show the best-fit phase curves obtained with the Hapke model using the coherent backscatter terms, coherent backscatter terms at $B_{C0_{\lambda}}$ = 0, and only shadow-hiding terms, respectively.
}
\label{fig:Hapke}
\end{figure}

\begin{figure}
  \begin{center}
    \FigureFile(80mm,80mm){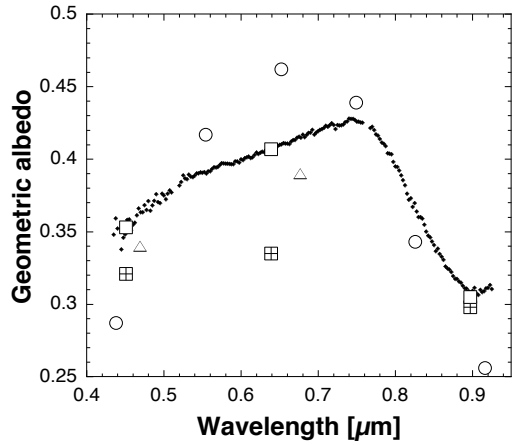}
  \end{center}
  \caption{
Comparison of geometric albedo for Vesta. 
The dots, open triangles, open circles, crossed squares, and open squares represent the geometric albedo from a spectroscopic study \citep{Bus2002}, HST \citep{Li2011}, Dawn \citep{Li2013}, and this study without and with coherent backscatter opposition effect terms, respectively. 
The Vestan spectrum is scaled to match the geometric albedo in the $R_{\rm C}$ band in this study with coherent backscatter opposition effect terms.
}
\label{fig:Hapke-spectra}
\end{figure}

The phase functions for Vesta in the $B$, $R_{\rm C}$, and $z'$ bands using the Hapke model, with or without the coherent backscatter opposition effect terms, are shown in Fig. \ref{fig:Hapke}.
$M_{\lambda}$(1, 1, 0) employing the coherent backscatter opposition effect terms that most closely fit the data gives $M_{B}$(1, 1, 0) = 3.83 $\pm$ 0.01, $M_{R_{\rm C}}$(1, 1, 0) = 2.67 $\pm$ 0.01, and $M_{z'}$(1, 1, 0) = 3.03 $\pm$ 0.01 mag.
Fig. \ref{fig:Hapke-spectra} shows the geometric albedo obtained in this and previous studies.
The spectral photometric data in this study, obtained using the coherent backscatter opposition effect terms, are in reasonably good agreement with the spectrum. 
A visible geometric albedo of 0.342 for Vesta by AKARI \citep{Usui2011} is consistent with this study without the coherent backscatter opposition effect term. 
This consistency occurs because the visible photometric data using AKARI did not include small phase angles.

\renewcommand{\tabcolsep}{1pt}
\begin{longtable}{ccccccccccccl}
  \caption{Comparison of the Hapke model for Vesta.}\label{tab:Hapke function}
  \hline
   $M_{\lambda}$(1, 1, 0) & $w_{\lambda}$ & $g_{\lambda}$ & $\bar{\theta_{\lambda}}$\footnotemark[$*$] &$B_{S0_{\lambda}}$ & $h_{S_{\lambda}}$ & $B_{C0_{\lambda}}$ & $h_{C_{\lambda}}$ & $A_{p_{\lambda}}$ & $q_{\lambda}$ & $A_{B_{\lambda}}$ &Band & References

\\ 
   $\rm[mag]$ & & & [$\timeform{D}$] & & [rad] & & [rad] & & & & &
\\
\endfirsthead
  \hline
   $M_{\lambda}$(1, 1, 0) & $w_{\lambda}$ & $g_{\lambda}$ & $\bar{\theta_{\lambda}}$ &$B_{S0_{\lambda}}$ & $h_{S_{\lambda}}$ & $B_{C0_{\lambda}}$ & $h_{C_{\lambda}}$ & $A_{p_{\lambda}}$ & $q_{\lambda}$ & $A_{B_{\lambda}}$ &Filter & References
\\
  \hline
\endhead
  \hline
\endfoot
\multicolumn{1}{@{}l}{\rlap{\parbox[t]{1.0\textwidth}{\small
\footnotemark[$*$]Assumed values were kept constant in the data fitting. \\
\footnotemark[$\dagger$]Fitting errors appear within brackets below each parameter value. \\
\footnotemark[$\ddagger$]See the third column in Table \ref{tab:slope parameter}.\\
\footnotemark[$\S$]The values were obtained from disk-resolved data of Vesta from approach and survey using color filters.\\
\footnotemark[$\|$]The solutions were obtained from the disk-integrated phase function of Vesta using approach data taken through a clear filter.\\
\footnotemark[$\#$]The values were obtained from photometric data using not only Vesta but also several V-type asteroids.
}}}
\endlastfoot
  \hline
\bf{3.83}&\bf{0.417}&\bf{-0.282}&\bf{(18.4)}&\bf{1.19}&\bf{0.052}&\bf{0.167}&\bf{0.0034}&\bf{0.353}&\bf{0.415}&\bf{0.146}&\boldmath $B$&\bf{This study}\\
$\langle0.01\rangle$&$\langle0.014\rangle$&$\langle0.035\rangle$&-----&$\langle0.20\rangle$&$\langle0.023\rangle$&$\langle0.130\rangle$&$\langle0.0060\rangle$&$\langle0.004\rangle$&$\langle0.004\rangle$&$\langle0.002\rangle$&&Fitting error\footnotemark[$\dagger$]\\
3.93&0.421&-0.310&(18.4)&1.10&0.032&-----&-----&0.321&0.450&0.145&$B$&This study\\
$\langle0.01\rangle$&$\langle0.012\rangle$&$\langle0.020\rangle$&-----&$\langle0.09\rangle$&$\langle0.006\rangle$&-----&-----&$\langle0.003\rangle$&$\langle0.005\rangle$&$\langle0.002\rangle$&&Fitting error\footnotemark[$\dagger$]\\
-----&0.393&-0.208&18.4&1.87&0.100&-----&-----&0.287&0.526&0.151&F8\footnotemark[$\ddagger$]&\citet{Li2013}\footnotemark[$\S$]\\
  \hline
\bf{2.67}&\bf{0.454}&\bf{-0.177}&\bf{(17.8)}&\bf{2.05}&\bf{0.086}&\bf{0.260}&\bf{0.0057}&\bf{0.407}&\bf{0.420}&\bf{0.171}&\boldmath $R_{\rm C}$&\bf{This study}\\
$\langle0.01\rangle$&$\langle0.020\rangle$&$\langle0.029\rangle$&-----&$\langle0.33\rangle$&$\langle0.028\rangle$&$\langle0.099\rangle$&$\langle0.0061\rangle$&$\langle0.004\rangle$&$\langle0.004\rangle$&$\langle0.002\rangle$&&Fitting error\footnotemark[$\dagger$]\\
2.88&0.464&-0.189&(17.8)&2.09&0.067&-----&-----&0.335&0.507&0.170&$R_{\rm C}$&This study\\
$\langle0.01\rangle$&$\langle0.011\rangle$&$\langle0.019\rangle$&-----&$\langle0.23\rangle$&$\langle0.011\rangle$&-----&-----&$\langle0.003\rangle$&$\langle0.005\rangle$&$\langle0.002\rangle$&&Fitting error\footnotemark[$\dagger$]\\
-----&0.556&-0.243&17.8&1.83&0.047&-----&-----&0.462&0.474&0.219&F7\footnotemark[$\ddagger$]&\citet{Li2013}\footnotemark[$\S$]\\
  \hline
\bf{3.03}&\bf{0.373}&\bf{-0.303}&\bf{(18.8)}&\bf{1.17}&\bf{0.040}&\bf{0.070}&\bf{0.0091}&\bf{0.305}&\bf{0.415}&\bf{0.127}&\boldmath $z'$&\bf{This study}\\
$\langle0.01\rangle$&$\langle0.011\rangle$&$\langle0.027\rangle$&-----&$\langle0.17\rangle$&$\langle0.026\rangle$&$\langle0.116\rangle$&$\langle0.0332\rangle$&$\langle0.003\rangle$&$\langle0.004\rangle$&$\langle0.001\rangle$&&Fitting error\footnotemark[$\dagger$]\\
3.06&0.374&-0.314&(18.8)&1.18&0.031&-----&-----&0.298&0.419&0.125&$z'$&This study\\
$\langle0.01\rangle$&$\langle0.009\rangle$&$\langle0.016\rangle$&-----&$\langle0.08\rangle$&$\langle0.005\rangle$&-----&-----&$\langle0.003\rangle$&$\langle0.004\rangle$&$\langle0.001\rangle$&&Fitting error\footnotemark[$\dagger$]\\
-----&0.365&-0.225&18.8&1.59&0.097&-----&-----&0.256&0.531&0.136&F4\footnotemark[$\ddagger$]&\citet{Li2013}\footnotemark[$\S$]\\
  \hline
-----&0.40&-0.30&(20)&1.03&0.044&-----&-----&-----&-----&-----&$V$&\citet{Helfenstein1989}\\
3.20&0.424&-0.150&(18.0)&2.60&0.120&-----&-----&0.329&0.538&0.177&$V$&\citet{Li2013}\footnotemark[$\|$]\\
-----&0.554&-0.244&17.5&1.83&0.048&-----&-----&0.417&0.468&0.195&F2\footnotemark[$\ddagger$]&\citet{Li2013}\footnotemark[$\S$]\\
-----&0.51&-0.26&31.5&1.0&0.098&-----&-----&0.34&0.44&0.15&$V$&\citet{Hicks2014}\footnotemark[$\#$]\\
  \hline
\end{longtable}

\subsection{Error analysis}
\label{sec:Error analysis}
In general, any standard statistical errors obtained from standard fitting routines including the Levenberg--Marquardt algorithm used in this study, premises that all parameters are independent of each other. 
However, it is known that each parameter of the Shevchenko function and Hapke models are strongly coupled (e.g., \cite{Belskaya2000}, \cite{Helfenstein2011}).  
The standard errors obtained from the algorithm underestimate uncertainties sometimes when applied to fits of the model which has coupled parameters. 
\citet{Helfenstein2011} indicated that standard statistical error estimation techniques can severely underestimate uncertainties in case of the model which has interdependence parameters such as the Hapke model. 
Therefore, the error for the Shevchenko function and Hapke models in this study may be underestimated.

\section{Discussion}
\label{sec:Discussion}
The magnitudes at a phase angle of 0$\timeform{D}$ ($M_{\lambda}$(1, 1, 0)) obtained with the Shevchenko function and Hapke model, including the coherent backscatter opposition effect terms, are $\sim$0.1 mag brighter than the absolute magnitude $H_{\lambda}$, which is equal to $M_{\lambda}$(1, 1, 0) obtained using the IAU $H$--$G$ function (Table \ref{tab:slope parameter}, \ref{tab:Shevchenko function}, and \ref{tab:Hapke function}). 
\citet{Belskaya2000} reported that approximations for various spectral types of asteroids using the $H$--$G$ function can deviate from the observed magnitude by up to 0.1 mag.
This reflects the fact that the $H$--$G$ function has only two parameters, which are insufficient to properly model the opposition effect.

\begin{figure}
  \begin{center}
    \FigureFile(80mm,80mm){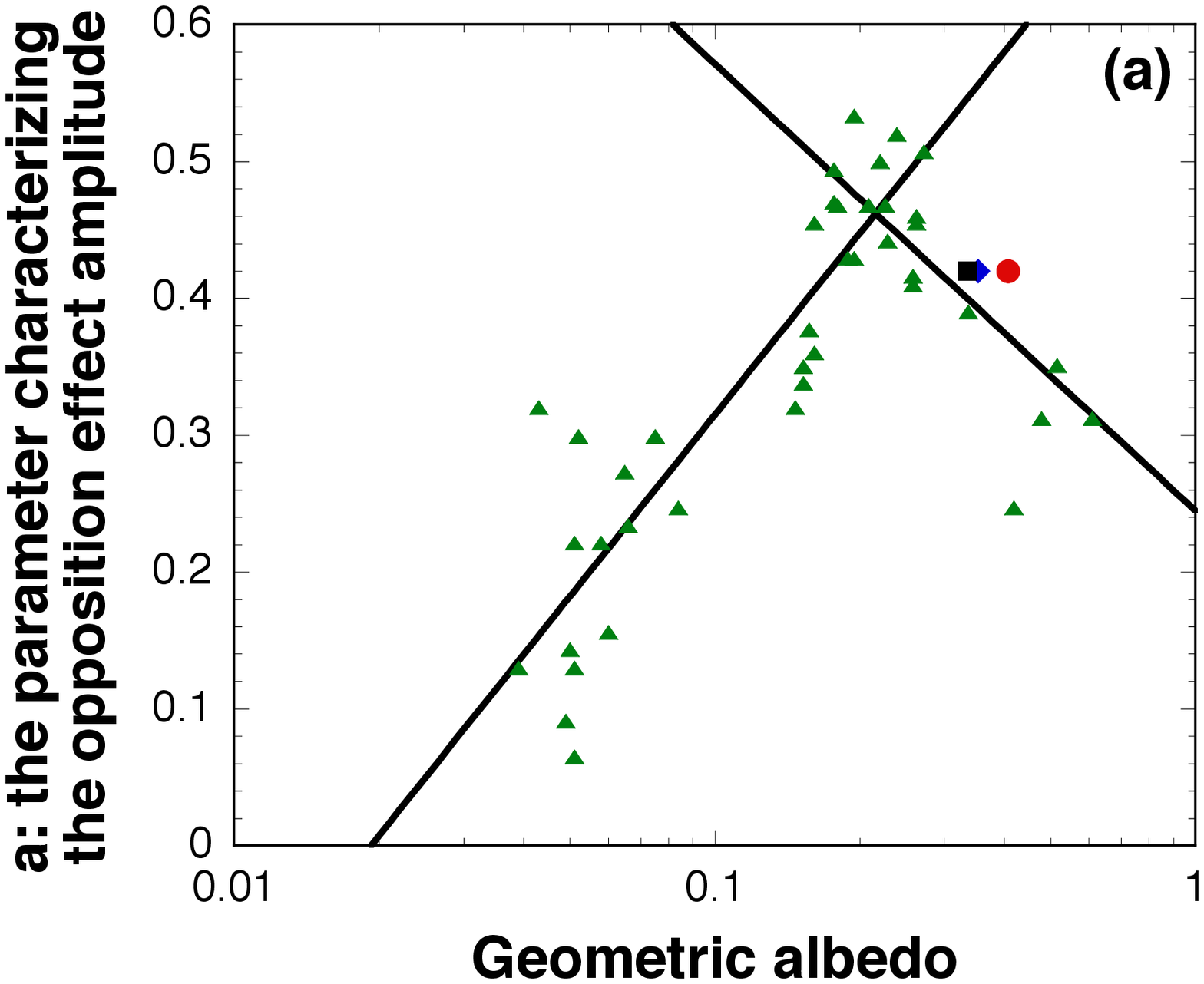}
    \FigureFile(80mm,80mm){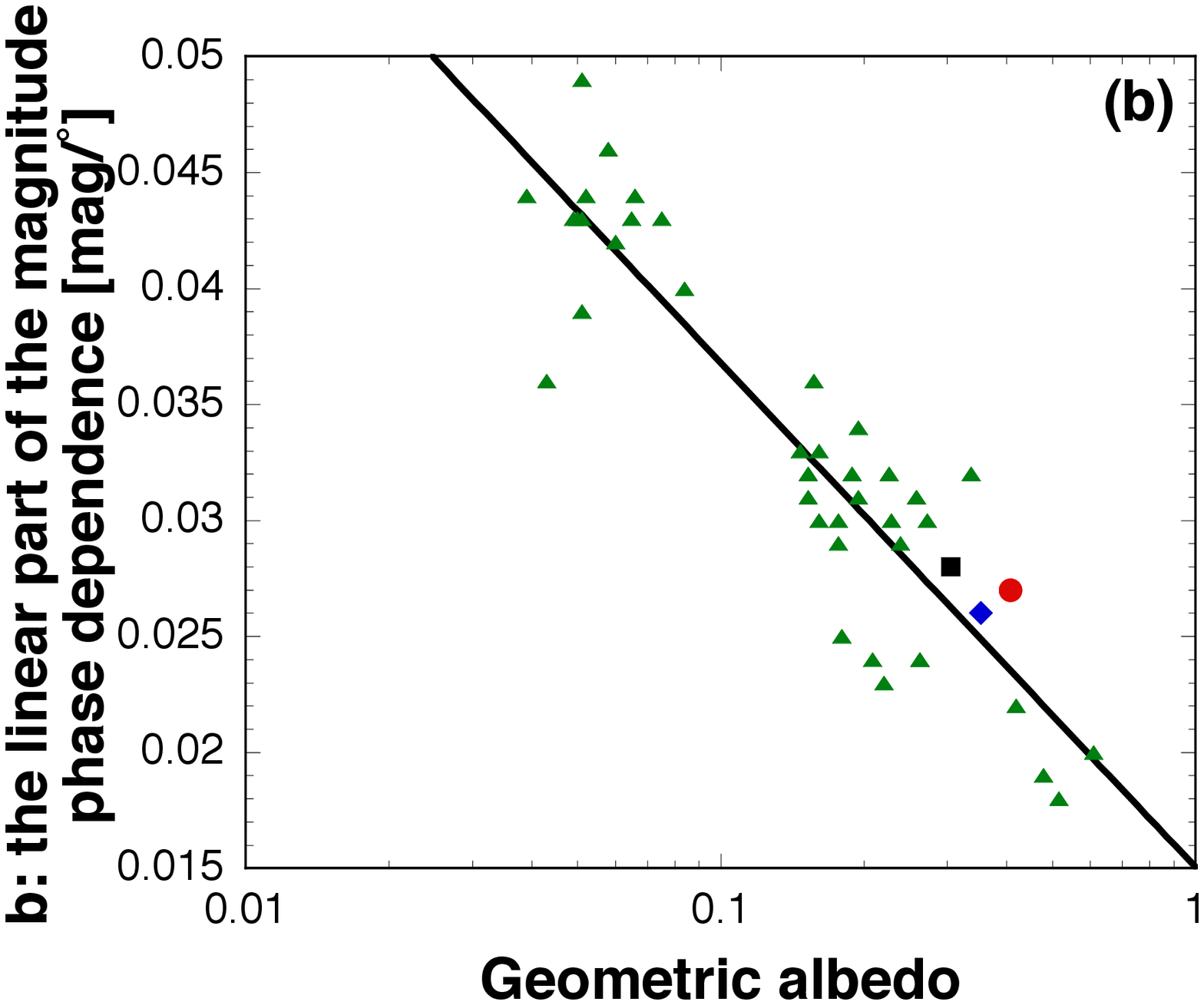}
  \end{center}
  \caption{Relationship between the geometric albedo and parameters of the Shevchenko function. 
The geometric albedo of Vesta was obtained from the Hapke model.
The parameters $a$ and $b$ for asteroids, apart from Vesta, were taken from \citet{Belskaya2002}, \citet{Shevchenko2002}, and \citet{Belskaya2003}.
The geometric albedos for other asteroids are from \citet{Usui2011}.
The blue diamond, red circle, black square, and green triangles are data for Vesta in the $B$, $R_{\rm C}$, and $z'$ band filters, and asteroids except Vesta, respectively. 
}
\label{fig:Shevchenko-ab}
\end{figure}

The Shevchenko function is a simple three-parameter model, which includes a parameter that directly expresses the opposition effect.
The distribution of the parameter describing the amplitude of opposition brightening ($a$) and the coefficient for the linear part of the phase curve ($b$) obtained by the Shevchenko model as a function of the geometric albedo of asteroids are shown in Fig. \ref{fig:Shevchenko-ab}.
\citet{Belskaya2000} showed that $a$ decreases for both dark and high-albedo asteroids, and that the largest $a$ values characterize moderate-albedo asteroids.
These authors also reported that $b$ increases linearly as the logarithm of geometric albedo decreases.
The Shevchenko function parameters for Vesta obtained in our study are not inconsistent with this previous study.
The correlation of parameters from the Shevchenko function can be expressed by:

\begin{equation} 
a = \left\{ \begin{array}{lcl}
0.24 - 0.33 \log{(p_{\lambda})}, 
& \mbox{for} & 0.22 < p_{\lambda}, \\
0.75 + 0.44 \log{(p_V)},
& \mbox{for} & p_{\lambda} < 0.22, \\
\end{array} \right. 
\label{eq:Shevchenko-a}
\end{equation}

\begin{eqnarray}
b = 0.015 - 0.022 \log{(p_{\lambda})},
\label{eq:Shevchenko-b}
\end{eqnarray}

\noindent
where $p_{\lambda}$ is the geometric albedo at the observed filter band. 
The geometric albedo can be estimated by the Shevchenko parameters in equation \ref{eq:Shevchenko-a} and/or \ref{eq:Shevchenko-b}.

\begin{figure}
  \begin{center}
    \FigureFile(80mm,80mm){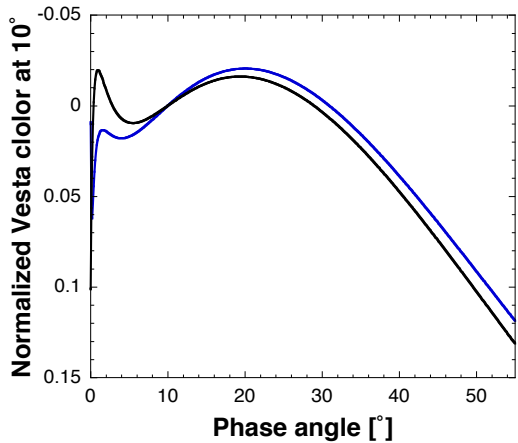}
  \end{center}
  \caption{
Vestan color as a function of phase angle using the Hapke model, including the coherent backscattering effect. 
The blue and black lines indicate ($B$ $\mathrm{-}$ $R_{\rm C}$) and ($z'$ $\mathrm{-}$ $R_{\rm C}$) for Vesta, respectively.
The color was normalized at a phase angle of 10$\timeform{D}$.
}
\label{fig:color from hapke}
\end{figure}

The spectra of Vesta in the wavelength range between the $B$ and $z'$ bands yield several-percent precision for phase angles from 8 to 53$\timeform{D}$ (see Fig. \ref{fig:spectrum}). 
The Vestan colors ($B$ $\mathrm{-}$ $R_{\rm C}$) and ($z'$ $\mathrm{-}$ $R_{\rm C}$) based on the Hapke model including the coherent backscattering effect (Fig. \ref{fig:color from hapke}) show differences in color within several percent up to 40$\timeform{D}$. This indicates a weak phase angle dependence of Vesta's spectrum.

The width of the shadow-hiding opposition effect peak $h_{S_{\lambda}}$ of the Hapke model can be interpreted in terms of the porosity and grain size distribution of the optically active regolith \citep{Hapke1986}.
$h_{S_{\lambda}}$ is given by:

\begin{eqnarray}
h_{S_{\lambda}} = - \frac{3}{8} Y \ln{\rho_{\lambda}},
\label{eq:hapke-h_S}
\end{eqnarray}

\noindent
where $Y$ and $\rho_{\lambda}$ describe the grain size distribution and porosity of the particles in an optically active scale, respectively.
Various cases are considered for $Y$ \citep{Hapke1986}.
$Y$ for the lunar regolith, which is probably similar to the surfaces of many other small bodies \citep{Hapke1986}, has the form:

\begin{eqnarray}
Y = \frac{\sqrt3}{\ln\left({\frac{d_l}{d_s}}\right)},
\label{eq:hapke-Y}
\end{eqnarray}

\noindent
where $d_l$ and $d_s$ describe the diameter of the largest and smallest particles of the optically active regolith, respectively.

The particle size of the Apollo soil samples ranges from several millimeters to several microns \citep{Carrier1973}.
Therefore, the ratio of $d_l$ and $d_s$ is $\sim 10^3$.
\citet{Gundlach2013} estimated the nature of the regolith of various bodies, including the Moon and Vesta, using remote measurements of the thermal inertia.
The calculated diameter of the lunar regolith material is $\sim$100 $\mu$m, which is in agreement with the mean value measured on Apollo samples.
The regolith grain size of Vesta is similar to that of the Moon.
The Dawn spacecraft has observed the surface of Vesta extensively. 
Its surface is heavily cratered, like that of the Moon (\cite{Jaumann2012}, \cite{Russell2012}).
The regolith of the Moon and Vesta formed as ejecta produced by collisional impacts at their surface (e.g., \cite{McKay1989}).
\citet{Hiroi1994} suggested that significant fine regolith particles (less than 25 $\mu$m) are present on the surface of Vesta.
These observations all imply that the $d_l$ and $d_s$ of Vesta are similar to those of the Moon.
Therefore, the ratio of $d_l$ and $d_s$ for the Moon is used in equation \ref{eq:hapke-Y} for Vesta.
The porosity of Vesta ($\rho_{B}$ = 0.575--0.7, $\rho_{R_{\rm C}}$ = 0.4--0.5, $\rho_{z'}$ = 0.65--0.7) is similar to that of the Moon \citep{Helfenstein1987}, and less than that of the asteroids 243 Ida and 433 Eros \citep{Domingue2002}, which have lower gravity than both Vesta and the Moon.
Most of the surface of Vesta is covered with eucrite-rich howardites and/or cumulate or polymict eucrites \citep{DeSanctis2013}. 
Howardites and eucrites have a grain density of between 3.0 $\times$ $10^3$ and 3.3 $\times$ $10^3$ kg $\mathrm{m^{-3}}$ \citep{Britt2003}. 
The bulk density is derived from (1 $\mathrm{-}$ porosity) $\times$ the grain density.
The bulk density of Vestan soil in the optically active region is 0.9--2.0 $\times$ $10^3$ kg $\mathrm{m^{-3}}$.
This is not inconsistent with a previously derived result that Vesta has a fairly loose and fluffy dust layer on its surface, based on passive microwave observations \citep{Redman1992}.

The bulk densities in each filter decrease with the reduction of the geometric albedos.
In general, the optical depth of the powder depends on the transparency of the particles.
The bulk density of lunar regolith increases with the depth \citep{Houston1974}.
Therefore, the density obtained at wavelengths of low reflectance may represent only information from shallower depths.

In this study, the results of the Hapke model with and without the coherent backscatter opposition effect terms are utilized.
The solution that minimizes $\chi^2$ for the model with the coherent backscatter opposition effect is smaller than that for the model without this effect.
The Shevchenko function parameter $a$ for Vesta identifies it as being on the high-albedo side, but near the peak (Fig \ref{fig:Shevchenko-ab}).
This implies that the opposition effect consists not only of the shadow-hiding effect, but also the coherent backscattering effect.

Fig. \ref{fig:Hapke}(b) highlights that the contribution of the coherent backscattering enhancement of Vesta appears at ca. 1.0$\timeform{D}$. 
\citet{Mishchenko2006} showed that the phase dependencies of the degree of linear polarization have a narrow local minimum of negative polarization centered at a phase angle that is approximately equal to a half-width of the corresponding coherent backscattering opposition effect.
\citet{Velichko2008} demonstrated that a phase dependence in polarization exists for Vesta within a range of phase angles from 0.6$\timeform{D}$ to 24.7$\timeform{D}$, and that a narrow local minimum of negative polarization does not appear at greater than 1.5$\timeform{D}$.
The aforementioned features indicate that coherent backscattering enhancement contributes to scattering of Vesta at less than 1.0$\timeform{D}$.

The width of the coherent backscattering opposition effect peak $h_{C_{\lambda}}$ from the Hapke model, which is interpreted in terms of the wavelength and transport mean free path in the medium \citep{Hapke2002}, is given by:

\begin{eqnarray}
h_{C_{\lambda}} = \frac{\lambda}{4 \pi \Lambda_{\lambda}},
\label{eq:hapke-h_C}
\end{eqnarray}

\noindent
where $\lambda$ is the wavelength and $\Lambda$ is the transport mean free path in the medium.

Table \ref{tab:transport} lists $h_{C_{\lambda}}$, $\lambda$, $\Lambda_{\lambda}$, and mean regolith particle sizes for asteroids, the Moon, and Jovian and Saturnian satellites.
$\Lambda_{\lambda}$ in the regolith of Vesta, the Moon, and asteroid 2867 Steins is approximately one and a half orders of magnitudes smaller than the mean particle size.
This can be explained, as the coherent backscattering opposition appears to be linked to the most tegmental particles.
Theoretical models of the coherent backscattering opposition effect predict that $h_{C_{\lambda}}$ should be proportional to the wavelength, but no wavelength dependence on width has been experimentally shown for the Moon \citep{Hapke2012}.
This fact is supported by weak dependence between wavelength and $h_{C_{\lambda}}$ for Enceladus and Phoebe. 
The value of $h_{C_{\lambda}}$ for Vesta seems to be affected by the wavelength of the observations.
Since $h_{C_{\lambda}}$ in this study is defined by only three points, it is really hard to say whether it is wavelength-dependent.

\begin{longtable}{lccccl}
  \caption{Comparison of the transport mean free path in a medium.}\label{tab:transport}
  \hline
   Object & $h_{C_{\lambda}}$  & $\lambda$  & $\Lambda_{\lambda}$ & regolith grain & References
\\ 
    & $\timeform{D}$ &  [$\mu$m] &  [$\mu$m] & size\footnotemark[$*$] [$\mu$m] & 
\\
\endfirsthead
  \hline
   Object & $h_{C_{\lambda}}$  & $\lambda$  & $\Lambda_{\lambda}$ & regolith grain & References
\\
  \hline
\endhead
  \hline
\endfoot
  \hline
\multicolumn{1}{@{}l}{\rlap{\parbox[t]{1.0\textwidth}{\small
\footnotemark[$*$] Data on regolith grain size were taken from \citet{Gundlach2013}. 
}}}
\endlastfoot
  \hline
4 Vesta&0.20&$B$&10.6&108&This study\\
4 Vesta&0.33&$R_{\rm C}$&~8.9&108&This study\\
4 Vesta&0.89&$z'$&~4.7&108&This study\\
  \hline
Enceladus&0.645&0.34&~2.4 &-----&\citet{Verbiscer2005}\\
Enceladus&0.17&0.55&14.6 &-----&\citet{Verbiscer2005}\\
Enceladus&0.28&0.87&14.1 &-----&\citet{Verbiscer2005}\\
  \hline
Europa&0.08&0.47&26.7 &-----&\citet{Verbiscer2005}\\
Ganymede&0.22&0.47&~9.8 &-----&\citet{Verbiscer2005}\\
2867 Steins&0.14 &0.63&20.1 &1260&\citet{Spjuth2012}\\
Moon&2.27&0.56&~1.1 &96&\citet{Verbiscer2005}\\
  \hline
Phoebe&0.64&$B$&~3.4 &-----&\citet{Miller2011}\\
Phoebe&0.81&$R$&~3.6 &-----&\citet{Miller2011}\\
Phoebe&0.82&$I$&~4.6 &-----&\citet{Miller2011}\\

\end{longtable}

\begin{figure}
  \begin{center}
    \FigureFile(80mm,80mm){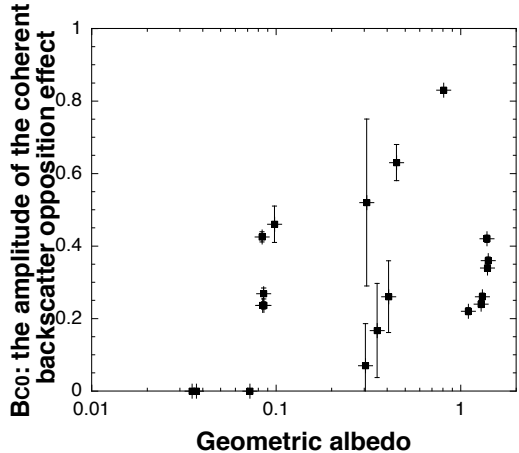}
  \end{center}
  \caption{Comparison of the strength $B_{C0}$ of the coherent backscattering term for solar system objects.
$B_{C0}$ values for the asteroids 588 Achilles, 884 Priamus, and 1143 Odysseus are taken as being zero, due to their lack of non-linear opposition brightening.
The values for Jovian Trojan asteroids, Steins, Moon, and Galilean satellites were taken from \citet{Usui2011}, \citet{Lamy2008}, \citet{Lane1973}, and \citet{Buratti1995}, respectively.
}
\label{fig:B_C0}
\end{figure}

\citet{Hapke2002} did not identify any relationship between the strength $B_{C0}$ of the coherent backscattering term and any other parameters, but \citet{Hillier1999}, who used the Hapke model with minor changes to the coherent backscatter term \citep{Helfenstein1997}, suggested that $B_{C0}$ is related to the fraction of light that is multiply scattered within a single particle and, hence, related to particle structure.
However, the relevance of $B_{C0}$ and other parameters has not been investigated using observational data.
Table \ref{tab:Hapke function} shows that $B_{C0}$ for Vesta is enhanced, as the reflectance of Vesta is high relative to the wavelength.
The spectral ratio of less than 1.0$\timeform{D}$ generally increases with longer wavelength \citep{Kaydash2013}.
Moreover, the intensity of the lunar spectra in \citet{Kaydash2013} appears to redden.
Thus, $B_{C0}$ for the Moon is also enhanced at increased reflectance.
Figure \ref{fig:B_C0} implies that $B_{C0_{\lambda}}$ for solar system objects tends to increase when surface reflectivity increases.
These facts suggest that a relationship exists between $B_{C0_{\lambda}}$ and reflectance.
\citet{Li2013} showed that about 20\%--30\% of scattered light on the surface of Vesta is multiply scattered.
\citet{Shevchenko2012} proposed that an absence of opposition effects for low-albedo objects can be interpreted to result from their very dark surfaces in which only single-light scattering is important.
Therefore, multiple-scattering on an optically active scale may contribute to the coherent backscattering term on the surface.  

\section{Summary}
$B$, $R_{\rm C}$, and $z'$ band photometric observations of Vesta were carried out, including measurements at very low phase angles. 
The magnitudes at a phase angle of 0$\timeform{D}$ in each band are $M_{B}$(1, 1, 0) = 3.83 $\pm$ 0.01, $M_{R_{\rm C}}$(1, 1, 0) = 2.67 $\pm$ 0.01, and $M_{z'}$(1, 1, 0) = 3.03 $\pm$ 0.01 mag.
The absolute magnitude obtained using the IAU $H$--$G$ function is $\sim$0.1 mag darker than that at the phase angle of 0$\timeform{D}$ determined from the Shevchenko function and Hapke models with the coherent backscattering effect term.
The Hapke model analysis of the obtained data resulted in a geometric albedo of 0.35 in the $B$ band, 0.41 in the $R_{\rm C}$ band, and 0.31 in the $z'$ bands.
The bulk density of the optically active regolith on Vesta was estimated based on the Hapke model at 0.9--2.0 $\times$ $10^3$ kg $\mathrm{m^{-3}}$.
The coherent backscattering effect contributes to the opposition effect on the surface of Vesta, at least at values below 1.0$\timeform{D}$.
The amplitude of the coherent backscatter opposition effect for solar system objects, apart from extreme high-albedo objects, increases as geometric albedo increases.
This supports the multiple-scattering interpretation of the coherent backscattering term on Vesta.

\bigskip
We thank the staff members of the Maidanak Astronomical Observatory for their support during photometric observations. 
We are grateful to Jian-Yang Li for sharing valuable details of Dawn spacecraft observations of Vesta. 
We thank an anonymous reviewer for their careful and constructive reviews, which helped improve the manuscript significantly.
This study was supported by the National Research Foundation of Korea (NRF) grant funded by the Korea Government (MEST) (No. 2012R1A4A1028713), Optical \& Near-Infrared Astronomy Inter-University Cooperation Program, the MEXT of Japan, and the Space Plasma Laboratory, ISAS, JAXA, as a collaborative research program.


\end{document}